\documentclass[sigconf,anonymous=false,10pt]{acmart}

\pdfoutput=1

\usepackage{booktabs} 
\usepackage{graphicx}
\usepackage{balance}  
\usepackage{amsmath}
\usepackage{color}

\def\BibTeX{{\rm B\kern-.05em{\sc i\kern-.025em b}\kern-.08emT\kern-.1667em\lower.7ex\hbox{E}\kern-.125emX}}

\definecolor{goodgreen}{rgb}{0.1, 0.5, 0.1}

\colorlet{dred}{red!80!black}
\colorlet{dgreen}{green!50!black}
\colorlet{dorange}{orange!90!black}
\colorlet{dblue}{blue!80!black}
\definecolor{oxfordblue}{rgb}{0, 0.33, 0.71}

\usepackage{url}

\usepackage[ruled, noend]{algorithm2e}

\usepackage{tikz,tikz-qtree,tikz-qtree-compat}
\usetikzlibrary{fit,topaths,calc,shapes,decorations,chains,scopes,arrows,calc,positioning}

\usepackage{subfigure}
\definecolor{goodgreen}{rgb}{0.1, 0.5, 0.1}

\usepackage[normalem]{ulem}
\usepackage{soul}

\newcommand{\pluseq}{\mathrel{+}=}

\newcommand{\nop}[1]{}

\newcommand{\inner}[1]{\left\langle #1 \right\rangle}

\newcommand{\mv}[1]{\mathbf{#1}}
\renewcommand{\vec}[1]{\ensuremath\boldsymbol{#1}}

\newcommand{\norm}[1]{\left\|#1\right\|}
\newcommand{\gv}[1]{\ensuremath{\mbox{\boldmath$ #1 $}}} 
\newcommand{\grad}[1]{\gv{\nabla} #1} 

\usepackage{todonotes}

\allowdisplaybreaks[1]

\begin{document}

\title{A Layered Aggregate Engine for Analytics Workloads}

\author{Maximilian Schleich}
\affiliation{\institution{University of Oxford}}
\author{Dan Olteanu}
\affiliation{\institution{University of Oxford}}
\author{Mahmoud Abo Khamis}
\affiliation{\institution{relational\underline{AI}}}
\author{Hung Q. Ngo}
\affiliation{\institution{relational\underline{AI}}}
\author{XuanLong Nguyen}
\affiliation{\institution{University of Michigan}}

\renewcommand{\shortauthors}{}

\setcopyright{none}
\renewcommand\footnotetextcopyrightpermission[1]{}
\settopmatter{printacmref=false}
\pagestyle{plain} 
\fancyhead{}

\begin{abstract}
This paper introduces LMFAO (Layered Multiple Functional Aggregate Optimization), an in-memory optimization and execution engine for batches of aggregates over the input database. The primary motivation for this work stems from the observation that for a variety of analytics over databases, their data-intensive tasks can be decomposed into group-by aggregates over the join of the input database relations. We exemplify the versatility and competitiveness of LMFAO for a handful of widely used analytics: learning ridge linear regression, classification trees, regression trees, and the structure of Bayesian networks using Chow-Liu trees; and data cubes used for exploration in data warehousing.

LMFAO consists of several layers of logical and code optimizations that systematically exploit sharing of computation, parallelism, and code specialization.

We conducted two types of performance benchmarks. In experiments with four datasets, LMFAO outperforms by several orders of magnitude on one hand, a commercial database system and MonetDB for computing batches of aggregates, and on the other hand, TensorFlow, Scikit, R, and AC/DC for learning a variety of models over databases.

\begin{quote}
{\em 
\hspace*{-1em}Aggregation is the aspirin to all problems.\\
\hspace*{4em} -- contemporary Greek philosopher
}
\end{quote}

\end{abstract}

\maketitle

\section{Introduction}
\label{sec:introduction}

This work has its root in two observations. First, the majority of practical analytics tasks involve relational data, with the banking or retail domains exceeding 80\%~\cite{Kaggle}. Second, for a variety of such analytics tasks, their data-intensive computation can be reformulated as batches of group-by aggregates over the join of the database relations~\cite{SOC:SIGMOD:16,ANNOS:DEEM:18}.

We introduce LMFAO (Layered Multiple Functional Aggregate Optimization), an in-memory optimization and execution engine for batches of aggregates over relational data. We exemplify the versatility and competitiveness of LMFAO for a handful of widely used analytics: learning ridge linear regression, classification trees, regression trees, and the structure of Bayesian networks using Chow-Liu trees; and data cubes used for exploration in data warehousing.

Query processing lies at the core of database research, with four decades of innovation and engineering on query engines for relational databases. Without doubt, the efficient computation of a handful of group-by aggregates over a join is well-supported by mature academic and commercial systems and also widely researched\nop{, e.g., the recent development on functional aggregate queries~\cite{faq} that rely on worst-case optimal join algorithms~\cite{NPRR12,LFTJ}}. There is relatively less development for large batches of such queries, with initial work in the context of data cubes~\cite{DataCube:ICDE:1996,DataCube:SIGMOD:1996,DataCube:SIGMOD:1997} and SQL-aware data mining systems~\cite{Chaudhuri:DMDB:1998,Classification:ICDE:1999} from two decades ago.

We show that by designing for the workload required by analytics tasks, LMFAO can outperform general-purpose mature database systems such as PostgreSQL, MonetDB, and a commercial database system by orders of magnitude. This is not only a matter of query optimization, but also of execution. Aspects of LMFAO's optimized execution for query batches can be cast in SQL and fed to a database system. Such SQL queries capture decomposition of aggregates into components that can be pushed past joins and shared across aggregates, and as such they may create additional intermediate aggregates. This poses scalability problems to these systems due to, e.g., design limitations such as the maximum number of columns or lack of efficient query batch processing, and led to larger compute times than for the plain unoptimized queries. This hints at LMFAO's distinct design that departs from mainstream query processing.

The performance advantage brought by LMFAO's design becomes even more apparent
for the end-to-end applications. For the aforementioned use cases in machine
learning, the application layer takes relatively insignificant time as it
offloads all data-intensive computation to LMFAO. LMFAO computes from the input
database sufficient statistics whose size ranges from tens of KBs to 
hundreds of MBs (Table~\ref{table:numberaggregates}) and that are used for learning
regression and classification models.  Mainstream solutions, e.g.,
MADlib~\cite{MADlib:2012}, R~\cite{R-project}, Scikit-learn~\cite{scikit2011},
\nop{XGBoost~\cite{xgboost},} and TensorFlow~\cite{tensorflow}, either take
orders of magnitude more time than LMFAO to train the same model or do not work
due to various design limitations. These solutions use data systems to
materialize the training dataset, which is defined by a feature extraction query
over a database of multiple relations, and ML libraries to learn models over
this dataset. We confirm experimentally that the main bottleneck of these
solutions is this materialization: The training datasets can be an order of
magnitude larger than the input databases used to create them
(Table~\ref{table:datasetstats}). In addition to being expected to work on much
larger inputs, the ML libraries are less scalable than the data systems.  Furthermore, these
solutions inherit the limitations of both of their underlying systems,
e.g., the maximum data frame size in R and the maximum number of columns in PostgreSQL are
much less than typical database sizes and respectively number of model features.

\subsection{Problem Statement}
\label{sec:problem}

LMFAO evaluates batches of queries of the following form:
\begin{align*}
&\texttt{SELECT } \hspace*{1em} F_1,\ldots, F_f, \texttt{ SUM}(\alpha_{1}), \ldots, \texttt{ SUM}(\alpha_{\ell})\\
&\texttt{FROM } \hspace*{2em} R_1 \texttt{ NATURAL JOIN } \ldots \texttt{ NATURAL JOIN } R_m\\
&\texttt{GROUP BY } F_1,\ldots, F_f;
\end{align*}
The user-defined aggregate functions (UDAFs), or simply aggregates, $\alpha_1,\ldots,\alpha_{\ell}$ can be sums of products of functions:

\begin{align*}
\forall i\in[\ell]:\hspace*{2em} \alpha_{i} = \sum_{j \in [s_i]} \prod_{k\in [p_{ij}]} f_{ijk}, \mbox{ where } s_i, p_{ij}\in\mathbb{N}
\end{align*}
We next give examples of such aggregates. To express count and sum aggregates, i.e., \texttt{SUM(1)} and \texttt{SUM}$(X_1)$ for some attribute $X_1$, we take $s_i=p_{ij}=1$ and then $f_{i11}$ is the constant function $f_{i11}()=1$ and respectively the identity function $f_{i11}(X_1)=X_1$. To encode a selection condition $X_1\text{ op } t$ that defines a decision tree node, where $X_1$ is an attribute, $\text{op}$ is a binary operator, and $t$ is a value in the domain of $X_1$, we use the Kronecker delta $f_{i11}(X_1) = \mathbf{1}_{X_1\text{ op } t}$, which evaluates to 1 if the condition is true and 0 otherwise. A further example is given by $s_i=n$, $p_{ij}=2$, and for $j\in[n]$ the constant functions $f_{ij1}() = \theta_j$ and the identity functions $f_{ij2}(X_j)=X_j$. Then, $\alpha_i$ is $\sum_{j\in[n]}\theta_j\cdot X_j$ and captures the linear regression function with parameters $\theta_j$ and features $X_j$. A final example is that of an exponential $n$-ary function $f_{i11}(X_1,\ldots,X_n) = e^{\sum_{j\in[n]}\theta_j\cdot X_j}$, which is used for logistic regression.

Applications, such as those in Section~\ref{sec:applications}, may generate batches of tens to thousands of aggregates (Table~\ref{table:numberaggregates}). They share the same join of database relations and possibly of relations defined by queries of the same above form.

\subsection{The Layers of LMFAO}

\begin{figure}[t]
  \begin{tikzpicture}[xscale=0.8, yscale=0.6]
    \tikzstyle{data} = [draw, rectangle, rounded corners = .07cm, align=center,
    inner sep = .2cm, outer sep = .1 cm]
    \tikzstyle{path} = [->, double, rounded corners=.1cm]
    
    \node[data,fill=dgreen!20,scale=0.8] at (4,-1.5) (app) {Application};
    \node[data,scale=0.8] at ($(app) + (0,-3)$) (cluster) {Aggregates};
    \node[data,scale=0.8] at ($(cluster) + (0,-3)$) (td) {Join Tree};

    
    \node[data,scale=0.8] at ($(td) + (4,0)$) (root)
    {Find Roots};

    \node[data,scale=0.8] at ($(root) + (0,3)$) (pushdown)
    {Aggregate\\ Pushdown};
    
    \node[data,scale=0.8] at ($(pushdown) + (0,3)$) (mergeview)
    {Merge Views};

    \node[data,scale=0.8] at ($(mergeview) + (4,0)$) (groupview) {Group Views};
    \node[data,scale=0.8] at ($(groupview) + (0,-2)$) (aggreg) {Multi-Output\\ Optimization};
    \node[data,scale=0.8] at ($(aggreg) + (0,-2)$) (parallel) {Parallelization};
    \node[data,scale=0.8] at ($(parallel) + (0,-2)$) (compilation) {Compilation};

    \node[scale=0.9] at (4,0.1)(n1) {\color{dred}App $\rightarrow$ LMFAO};
    \node[scale=0.9] at (8,0.1)(n1) {\color{dred} Logical Optimization};
    \node[scale=0.9] at (12,0.1)(n1) {\color{dred} Code Optimization};    
    
    \draw[dashed,thick,red] (5.8,0.5)--(5.8,-8.5);
    \draw[dashed,thick,red] (10,0.5)--(10,-8.5);

    \draw[path] (app) -- (cluster);
    \draw[path] (cluster) -- (td);

    \draw[path] (td) -- ($(td)+(0,-1.1)$) -- ($(root)+(0,-1.1)$) -- (root);
    \draw[path] (root) -- (pushdown);
    \draw[path] (pushdown) -- (mergeview);

    \draw[path] (mergeview) -- ($(mergeview)+(0,1.1)$) -- ($(groupview)+(0,1.1)$) --
    (groupview);

    \draw[path] (groupview) -- (aggreg);
    \draw[path] (aggreg) -- (parallel);
    \draw[path] (parallel) -- (compilation);
   
  \end{tikzpicture}
  \caption{The Optimization Layers of LMFAO.}
  \label{fig:lmfao-architecture}
  \vspace*{-1em}
\end{figure}
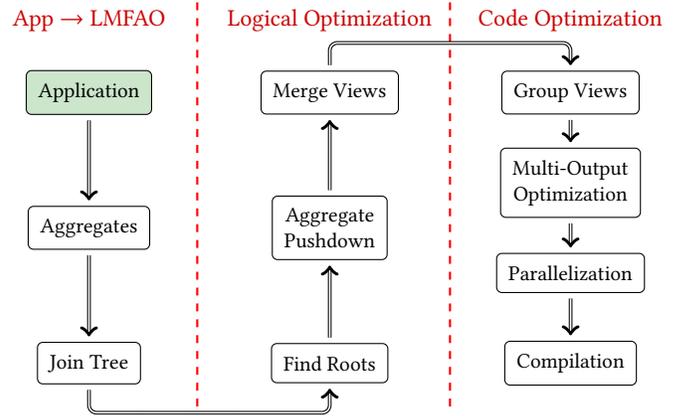

To evaluate aggregate batches, LMFAO employs a host of techniques, either novel or adaptations of known concepts to our specific workload. The layered architecture of LMFAO is given in Figure~\ref{fig:lmfao-architecture} and highlighted next. Section~\ref{sec:engine} expands on the key design choices behind LMFAO.

The \fbox{Join Tree} layer takes as input the batch of aggregates, the database
schema, and cardinality constraints (e.g., sizes of relations and attribute
domains) and produces one join tree that is used to compute all aggregates. This
step uses state-of-the-art techniques\footnote{For cyclic queries, we
  first compute a hypertree decomposition and materialize its bags (cycles) to
  obtain a join tree.}~\cite{abiteboul1995foundations}.

\nop{
  This step uses known state-of-the-art techniques. We first construct a
  hypertree decomposition of the underlying natural join whose width matches the
  fractional hypertree width of the join enhanced with cardinality
  constraints~\cite{Marx:2010,panda17}, and then materialize its bags (cycles)
  using a worst-case optimal join algorithm~\cite{NPRR12,LFTJ}. The outcome is a
  tree where each node is a (input or computed) relation. In the common case
  where the join is acyclic, then the decomposition step is not needed. Prior
  work showed that to achieve the lowest known complexity for processing
  aggregates with different group-by clauses, it is crucial to use different
  join trees, where the nodes containing attributes in the group-by clause form
  a connected subtree~\cite{BKOZ13,faq}. We found this highly impractical as it
  requires too many recomputations of (parts of) the join and precludes sharing
  across the aggregates.  }

The \fbox{Find Roots} layer is novel and affects the design of all subsequent
layers. By default, LMFAO computes each group-by aggregate in one bottom-up pass
over the join tree, by decomposing the aggregate into views computed along each
edge in the join tree. We allow for different traversals of the join tree:
different aggregates may be computed over the same join tree rooted at different
nodes. This can reduce the overall compute time for the batch as it can reduce
the number of views and increase the sharing of their computation. In our
experiments, the use of multiple roots for the computation of aggregate batches
led to $2 - 5\times$ speedup.

LMFAO uses {\em directional views} to support different traversals of the join tree: For each edge between two nodes, there may be views flowing in both directions. Directional views are similar in spirit with messages in the message passing algorithm used for inference in graphical models~\cite{Pearl:MessagePassing:1982}. Figure~\ref{fig:join-tree-favorita}~(middle) depicts directional views along a join tree.

In the \fbox{Aggregate Pushdown} layer, each aggregate is decomposed into one directional view per edge in the join tree. Its view at an edge going out of a node $n$ computes the aggregate when restricted to the subtree rooted at $n$ and is defined over the join of the views at the incoming edges of $n$ and of the relation at $n$. The directions of these views are from the leaves to the root of the aggregate. The rationale for this decomposition is twofold. First, it partially pushes the aggregates past joins (represented by edges in the tree), as in prior work on eager computation of one aggregate~\cite{Larson:LazyEager:1995} and its generalization to factorized databases~\cite{BKOZ13}. Second, it allows for sharing common views across the aggregates. 

The \fbox{Merge Views} layer consolidates the views generated in the previous layer. There are three types of merging possible for views along the same edge in the join tree, depending on whether they have the same: group-by attributes; aggregates; and body. Views with the same direction are defined over the same subtree of the join tree. We first identify identical views constructed for different aggregates and only keep one copy. We then merge the views with the same group-by attributes and body but different aggregates. We finally merge views with the same group-by attributes and different bodies. This consolidation is beneficial. For instance, there are 814 aggregates to compute for learning a linear regression model over the join of five relations in our Retailer dataset. This amounts to $814 \text{ aggregates}\times 4 \text{ edges } = 3,256 \text{ views}$, which are consolidated into $34$ views that have between themselves 1,468 aggregates.

The previous three layers are concerned with logical transformations of view expressions. The remaining layers consider optimizations not expressible at the syntactic level. 

In the \fbox{Group Views} layer, we group the views going out of the same node possibly along different edges such that there is no dependency between them. No dependency means that they can be evaluated together once the incoming views used in their joins are computed. The views in a group do not necessarily have the same group-by attributes, so a view group has multiple outputs. To continue our example, the remaining $34$ views are clustered into $7$ groups. 

The view group is a computational unit in LMFAO. At the \fbox{Multi-Output Optimization} layer, we construct the execution plan for each view group at a node. This plan needs one pass over the relation at that node, with lookups using the join keys into the incoming views to fetch aggregates needed for the computation of the views in the group. This is yet another instance of sharing in LMFAO: The computation of different views share the scan of the relation at the node. This is particularly beneficial for snowflake schemas with large fact relations, e.g., Inventory in Retailer and Sales in Favorita datasets. This scan sees the relation organized logically as a trie: first grouped by one attribute, then by the next in the context of values for the first, and so on. This trie organization is reminiscent of factorized databases~\cite{BaOlZa12} and LeapFrog TrieJoin~\cite{LFTJ} and can visit up to three times less values than a standard row-based scan for our datasets. In our experiments, this layer brought $1.4-2\times$ extra speedup.

The \fbox{Parallelization} layer addresses task and domain parallelism. LMFAO parallelizes the computation of multi-output plans for view groups that do not depend on each other. For this, it computes the dependency graph of the view groups. LMFAO partitions the largest input relations and allocates a thread per partition to compute the multi-output plan on that partition. This layer brought $1.4-3\times$ extra speedup on a machine with four vCPUs (AWS d2.xlarge).

Finally, the \fbox{Compilation} layer generates C++ code for the parallel
execution of multi-output plans. This code is specialized to the join tree and
database schema, with separate code for each view group and also for general
tasks such as data loading. The separate code chunks are compiled in
parallel.  The code layout for each view group is designed to maximize the computation sharing
across many aggregates with different group-by and UDAFs via the introduction of local variables, and to minimize the
number of accesses (initialization, update, lookup) to these local variables. 
LMFAO adopts various low-level code optimizations: inlining function calls; organization of the
aggregates for each view in a contiguous fixed-size array and ordered to allow
sequential read/write; reuse of arithmetic operations, e.g., repeating
multiplication of entries in the aggregate array; and synthesis of loops
from long sequences of lockstep computations. The latter two
optimizations are enabled by sorted input relations and views that are accessed
in lockstep. The organization of aggregates allows us to manage them in
contiguous batches. This is reminiscent of vectorization~\cite{Vectorwise:ICDE:2012}, now
applied to aggregates instead of data records.



\nop{
  The specialized code chunks are compiled in parallel. LMFAO uses an
  intermediate representation for this code and applies various optimizations,
  including: inlining function calls; organization of aggregate computation to
  minimize the number of updates to each aggregate; organization of the aggregates
  for each view in a contiguous fixed-size array and ordered to allow sequential
  read/write; reuse of arithmetic operations, such as repeating multiplication of
  two entries in the aggregate array; and synthesis of loops from long sequences
  of lockstep aggregate computations. The latter two optimizations are enabled by
  sorted input relations and views that are accessed in lockstep. The organization
  of aggregates allows us to manage them in batch. This is reminiscent of
  vectorization~\cite{Vectorwise:ICDE:2012}, now applied to aggregates instead of
  data records. LMFAO generates C++ code out of the optimized intermediate
  representation.
}

Some applications require the computation of UDAFs that change between iterations depending on the outcome of computation. For instance, the nodes in a decision tree are iteratively constructed in the context of conditions that are selected based on the data. The application tags these functions as {\em dynamic} to instruct LMFAO to avoid inlining their calls and instead generate separate code that is compiled between iterations and linked dynamically.

\subsection{Contributions}

To sum up, the contributions of this work are as follows:

1. We introduce LMFAO, a principled layered approach to computing large batches of group-by aggregates over joins. Its layers encompass several stages of logical and code optimization that come with novel contributions as well as adaptations of known techniques to a novel setting. The novel contributions are on: using different traversals of the same join tree to solve many aggregates with different group-by clauses; synthesizing directional views out of large sets of views representing components of aggregate queries; and the multi-output execution plans for computing groups of directional views using one pass over the input relations. It adapts compilation techniques to generate specialized code for the parallel computation of multi-output plans for aggregate queries with static and dynamic user-defined functions. 

2. We show the versatility of LMFAO for a range of analytics applications built on top of it. 

3. We implemented LMFAO in C++ and conducted two kinds of performance
benchmarks: The computation of aggregate batches and of end-to-end applications
using these aggregates. In experiments with four datasets, LMFAO outperforms
by several orders of magnitude on one hand, PostgreSQL, MonetDB and a commercial
DBMS for computing aggregate batches, and, on the other hand, TensorFlow,
Scikit, R, and AC/DC for learning models over databases.

\nop{

A taxonomy of existing data systems for learning models is given in Section~\ref{sec:relatedwork}.

tension between pipelining along the entire join tree and sharing; we go for sharing as it brings more extensive performance benefits here.

}

\section{Applications}
\label{sec:applications}

LMFAO encompasses a unified formulation and processing of core data processing tasks in database, data mining, and machine learning problems. We exemplify with a small sample of such problems: data cubes; gradients and covariance matrices used for linear regression, polynomial  regression, factorization machines; classification and regression trees; and mutual information of pairwise variables used for learning the structure of Bayesian networks.

We next introduce a compact query syntax and use it to formulate the above-mentioned data processing tasks.

\subsubsection*{Query Language}

We are given a database $D$ of $m$ (materialized) relations $R_1,\ldots,R_m$ over relation schemas $\omega_{R_1},\ldots,\omega_{R_m}$. For convenience, we see relation schemas, which are lists of attributes, also as sets. The list of attributes in the database is denoted by $\mathbf{X}=\bigcup_{r\in[m]}\omega_{R_r}=(X_1,\ldots,X_n,X_{n+1})$. 

We would like to compute a set of group-by aggregates over the natural join of these relations. This join may represent the training dataset for machine learning models, the input for multi-dimensional data cubes, or the joint probability distribution to be approximated by a Bayesian network. 

We use the following query formulation, which is more compact than the SQL form from Section~\ref{sec:problem}:
    \begin{align}
      Q(F_1,\ldots,F_f; \alpha_{1}, \ldots, \alpha_{\ell}) & \text{ += } R_1(\omega_{R_1}), \ldots, R_m(\omega_{R_m}) 
      \label{eq:query}
    \end{align}
In the head of $Q$, the group-by attributes $F_1,\ldots,F_f$ are separated from the aggregate functions by semicolon; we omit the semicolon if there are no group-by attributes. The aggregate functions are as defined in Section~\ref{sec:problem}. We use += to capture the SUM over each aggregate function. In the query body, we make explicit the attributes of each relation for a clearer understanding of the definitions of the aggregate functions. By definition, there is a functional dependency $F_1,\ldots,F_f\rightarrow \alpha_{1}, \ldots, \alpha_{\ell}$.

Our queries generalize FAQ-SS~\cite{faq} and MPF (Marginalize a Product Function)~\cite{MPF:2000} by allowing tuples of arbitrary UDAFs.


\subsubsection*{Ridge Linear Regression}
Assuming one parameter $\theta_j$ per attribute (feature) $X_j$, the linear
regression model is given by:
\begin{small}
  \begin{align*}
    LR(X_1,\ldots,X_n) &= \sum\nolimits_{j \in [n]} \theta_j\cdot X_j
  \end{align*}
\end{small}
In practice, features may be defined by attributes in both the input relations and results of queries over these relations. 

We assume without loss of generality that (1) $X_1$ only takes value $1$ and then $\theta_1$ is the so-called intercept and (2) $X_{n+1}$ is the label and has a corresponding new parameter  $\theta_{n+1}=-1$.

The error of the model is given by an objective function that is the sum of the least squares loss function and of the penalty term that is the $\ell_2$ norm of the parameter vector $\vec\theta$:
\begin{small}
  \begin{align*}
    J(\vec\theta)
    &= \frac{1}{2|D|}  \sum_{\mathbf{X}\in D} \big(
    \sum_{j \in [n]} \theta_j\cdot X_j - X_{n+1} \big)^2 + \frac{\lambda}{2} \norm{\vec\theta}^2\\
    &= \frac{1}{2|D|}  \sum_{\mathbf{X}\in D} \big(
    \sum_{j \in [n+1]} \theta_j\cdot X_j \big)^2 +
    \frac{\lambda}{2} \norm{\vec\theta}^2
  \end{align*}
\end{small}
We optimize the model using batch gradient descent (BGD), which updates the parameters in the direction of the gradient vector $\grad J(\vec\theta)$ of $J(\vec\theta)$ using a step size $s$:
\begin{small}
      \begin{align*}
      \texttt{repeat unti}&\texttt{l convergence:}\\
        \forall k \in [n] : \theta_k
        &:= \theta_k - s\cdot\grad_k J(\vec\theta) \\
        &:= \theta_k - s\cdot (\frac{1}{|D|} \sum_{\mv X\in D}
          (\sum_{j \in [n+1]} \theta_j\cdot X_j)\cdot X_k + \lambda \theta_k)
      \end{align*}
\end{small}
The above update relies on the aggregates for the size of the dataset $D$ and
the product of $X_k$ with the inner product
$\inner{\vec\theta,\mv X}=\sum_{j \in [n+1]} \theta_j\cdot X_j$. There are two
ways to express these aggregates. The common approach, which we call the
gradient vector, is to compute this inner product and then, for each gradient
$k$, multiply it with the corresponding $X_k$. This requires recomputation for
each new vector of parameters. The second approach~\cite{SOC:SIGMOD:16} is to
rewrite $\sum_{\mv X\in D}(\sum_{j \in [n+1]} \theta_j\cdot X_j)\cdot X_k$ as
$\sum_{j \in [n+1]} \theta_j\cdot \sum_{\mv X\in D}(X_j\cdot X_k)$ and compute
the non-centered covariance matrix (the covar matrix hereafter).

The covar matrix accounts for all pairwise multiplications $X_j\cdot X_k$. Each entry
  can be computed as aggregate query:
    \begin{small}
    \begin{align}
      Covar_{j,k}&(X_j \cdot X_k) \text{ += } R_1(\omega_{R_1}), \ldots, R_m(\omega_{R_m}). \label{eq:lr-sigma}
    \end{align}
  \end{small}
  Categorical attributes are one-hot encoded in a linear regression model. In
  our formalism, such attributes become group-by attributes. If only $X_j$ is
  categorical, we get:
  \begin{small}
    \begin{align}
      Covar_{j,k}&(X_j; X_k) \text{ += } R_1(\omega_{R_1}), \ldots, R_m(\omega_{R_m}) 
    \end{align}
  \end{small}
  If both $X_j$ and $X_k$ are categorical, we get instead:
  \begin{small}
    \begin{align}
      Covar_{j,k}&(X_j, X_k; 1) \text{ += } R_1(\omega_{R_1}), \ldots, R_m(\omega_{R_m}) 
    \end{align}
  \end{small}
  The computation of the covar matrix does not depend on the parameters $\vec\theta$, and
  can be done once for all BGD iterations.

\subsubsection*{Higher-degree Regression Models} A polynomial regression models of degree $d$ is defined as follows:
\begin{small}
\begin{align*}
PR_d(X_1&,\ldots,X_n) = \sum_{(a_1,\dots,a_{n})\in\mv A} \theta_{(a_1,\dots,a_{n})}\cdot \prod_{j=1}^{n} X_j^{a_j}, \mbox{ where }\\
\mv A &=\{(a_1,\ldots,a_n)\mid a_1+\ldots+a_n\leq d, \forall j\in[n]: a_j \in\mathbb{N}\}
\end{align*}
\end{small}
The covar matrix for $PR_d$ has the following aggregates in the gradient of the square loss function:
\begin{small}
\begin{align}
 &Covar_{(a_1,\dots,a_{n+1})} (X_1^{a_1}\cdot\ldots\cdot X_{n+1}^{a_{n+1}}) \text{ += } R_1(\omega_{R_1}), \ldots, R_m(\omega_{R_m}) \nonumber\\
 &\forall (a_1,\dots,a_{n+1}): \sum_{j=1}^{n+1} a_j \leq 2d, a_{n+1} \leq 1, \forall j\in[n+1]: a_j \in\mathbb{N}\label{eq:pr-sigma}
 \end{align}
\end{small}
A similar generalization works for factorization machines \cite{OS:SIGREC:2016,ANNOS:PODS:2018}. Categorical attributes can be accommodated as for linear regression and then each categorical attribute $X_j$ with exponent $a_j>0$ becomes a group-by attribute.

\subsubsection*{Data Cubes}

Data cubes~\cite{DataCube:ICDE:1996} are popular in data warehousing scenarios.  For a set $S_k\subseteq\mathbf{X}$ of $k$ attributes or dimensions, a $k$-dimensional data cube is a shorthand for the union of $2^k$ cube aggregates with the same aggregation function $\alpha$ over the same (measure) attribute out of $v$ attributes. We define one aggregate for each of the $2^k$ possible subsets of $S_k$:
\begin{small}
    \begin{align}
      \forall F_i\subseteq S_k:\ Cube_i(F_i; \alpha_1,\ldots,\alpha_v)
      &\text{ += } R_1(\omega_{R_1}), \ldots, R_m(\omega_{R_m}) \label{eq:data-cube}
    \end{align}
\end{small}
The cube aggregates have a similar structure with covar matrices for polynomial regression models. They both represent sets of group-by aggregates over the same join. However, the two constructs compute different aggregates and use different data representations. Whereas all cube aggregates use the same measure aggregation, the covar aggregates sum over different attributes. Data cubes are represented as tables in 1NF using a special ALL value standing for a set of values, whereas the covar matrices for regression models are matrices whose entries are the regression aggregates whose outputs have varying sizes and arities. 

A polynomial regression model of degree $d$ (PR$_d$) over categorical features given by $k$ attributes ($k \geq 2d$) requires regression aggregates whose group-by clauses are over all subsets of size at most $2d$ of the set of $k$ attributes. In contrast, a $2d$-dimensional data cube for a given set of $2d$ (dimension) attributes defines aggregates whose group-by clauses are over all subsets of the $2d$ attributes. The set of group-by clauses used by the aggregates for PR$_d$ is captured by all $2d$-dimensional data cubes constructed using the $k$ attributes.

\subsubsection*{Mutual Information} The mutual information of two distinct discrete random variables $X_i$ and $X_j$ is a measure of their mutual dependence and determines how similar the joint distribution is to the factored marginal distribution.  In our database setting, we capture the distributions of two attributes $X_i$ and $X_j$ using the following count queries that group by any subset of $\{X_i,X_j\}$ (thus expressible as a 2-dimensional data cube with a count measure):
\begin{small}
\begin{align}
 \forall S\subseteq\{i,j\}: Q_S((X_k)_{k\in S}; 1) &\text{ += } R_1(\omega_{R_1}), \ldots, R_m(\omega_{R_m}) \label{eq:mi}
\end{align}
\end{small}
The mutual information of $X_i$ and $X_j$ is the given by the following query with a $4$-ary aggregate function $f$ over the aggregates of the queries $Q_S$ defined above:
\begin{small}
\begin{align*}
MI(f(\alpha,\beta,\gamma,\delta)) &\text{+=} Q_\emptyset(\alpha), Q_{\{i\}}(X_i; \beta), Q_{\{j\}}(X_j; \gamma), Q_{\{i,j\}}(X_i, X_j; \delta)\\
f(\alpha,\beta,\gamma,\delta) &= \frac{\delta}{\alpha} \cdot \log \left( \frac{\alpha \cdot \delta}{\beta \cdot \gamma}\right)
\end{align*}
\end{small}
Mutual information has many applications as it is used: as cost function in learning decision trees; in determining the similarity of two different clusterings of a dataset; as criterion for feature selection; in learning the structure of Bayesian networks. The Chow-Liu algorithm~\cite{Chow-Liu-trees:1968} constructs an optimal tree-shaped Bayesian network $T$ with one node for each input attribute in the set $\mathbf{X}$. It proceeds in rounds and in each round it adds to $T$ an edge $(X_i,X_j)$ between the nodes $X_i$ and $X_j$ such that the mutual information of $X_i$ and $X_j$ is maximal among all pairs of attributes not chosen yet.

\subsubsection*{Classification and Regression Trees} 
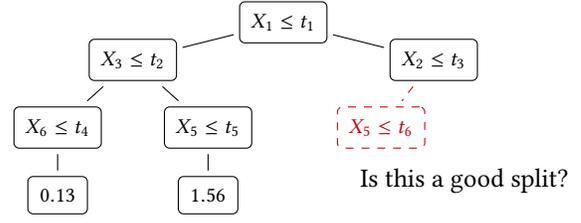
\begin{figure}[t]\centering
  \begin{tikzpicture}[yscale = 1, xscale=1, every node/.style={transform shape}]
    \tikzstyle{data} = [draw, rectangle, rounded corners = .07cm, align=center,
    inner sep = .2cm, outer sep = .1 cm]
        
    \node[data,scale=.8] (n1) {$X_1 \leq t_1$};
    \node[data,scale=.8] at($(n1)+(-2,-0.5)$) (n2) {$X_3 \leq t_2$};
    \node[data,scale=.8] at($(n1)+(2,-0.5)$) (n3) {$X_2 \leq t_3$};
    \node[data,scale=.8] at($(n2)+(-1,-0.9)$) (n4) {$X_6 \leq t_4$};
    \node[data,scale=.8] at($(n2)+(1,-0.9)$) (n5) {$X_5 \leq t_5$};
    \node[data,scale=.8,dashed,color=dred] at($(n3)+(-.7,-0.9)$) (n6) {$X_5 \leq t_6$};
    \node[data,scale=.8] at($(n4)+(0,-.9)$) (r1) {$0.13$};
    \node[data,scale=.8] at($(n5)+(0,-.9)$) (r2) {$1.56$};
    \node[anchor=west] at($(n6)+(-0.4,-.7)$) (r3) {Is this a good split?};
    
    \draw (n1) -- (n2); 
    \draw (n1) -- (n3); 
    \draw (n2) -- (n4); 
    \draw (n2) -- (n5);
    \draw (n4) -- (r1); 
    \draw (n5) -- (r2);
    \draw[dashed,color=dred] (n3) -- (n6);
  \end{tikzpicture}
  \caption{Example of a regression tree. Classification trees replace numerical
    leaves by categorical values.}
  \label{fig:decision-tree}
  \vspace*{-1em}
\end{figure}

Decision trees are popular
machine learning models that use trees with inner nodes representing conditional
control statements to model decisions and their consequences. Leaf nodes
represent predictions for the label. If the label is continuous, we learn a
regression tree and the prediction is the average of the label values in the
fragment of the training dataset that satisfies all control statements on the
root to leaf path.  If the label is categorical, the tree is a classification
tree, and the prediction is the most likely category for the label in the
dataset fragment.  Figure~\ref{fig:decision-tree} shows an example of a
regression tree.

The CART algorithm~\cite{cart84} constructs the tree one node at a time. Given
an input dataset $D$, CART repeatedly finds a condition $X_j \text{ op } t$ on
one of the attributes $X_1,\ldots,X_n$ of $D$ that splits $D$ so that a given
cost function over the label $X_{n+1}$ is minimized. For categorical attributes
(e.g., city), $t$ may be a set of categories and $\text{op}$ denotes
inclusion. For continuous attributes (e.g., age), $t$ is a real number and
$\text{op}$ is inequality.  Once this condition is found, a new node
\fbox{$X_j\text{ op } t$} is constructed and the algorithm proceeds recursively
to construct the subtree rooted at this node for the dataset representing the
fragment of $D$ satisfying the conditions at the new node and at its ancestors.

Practical implementations of CART compute at each node the cost for 20-100
conditions per continuous attribute and for categorical attributes the best
subset of categories is chosen based on the cost of splitting on each individual
category.

For regression trees, the cost is given by the variance:
\begin{small}
    \begin{align*}
      \text{variance}
      &= \sum_{\mathbf{X} \in D_i} X_{n+1}^2 -
        \frac{1}{|D_i|} \left(\sum_{\mathbf{X} \in D_i} X_{n+1}\right)^2
    \end{align*}
\end{small}
It is computed over the fragment $D_i$ of the dataset $D$. For the
tree depicted in Figure~\ref{fig:decision-tree},
$D_i = \sigma_{X_1 \geq t_1 \land X_2 \leq t_3 \land X_5 \leq t_6}(D)$, where
$X_5 \leq t_6$ is the new condition for which we compute the cost of the split
in the context of the conjunction of conditions
$X_1 \geq t_1 \land X_2 \leq t_3$ representing its ancestors in the tree.  The
computation of this cost needs the aggregates {\tt COUNT()}, {\tt
  SUM($X_{n+1}$)}, and {\tt SUM($X_{n+1}^2$)} over $D_i$:
\begin{small}
  \begin{align}
    RT (1 \cdot \alpha, X_{n+1} \cdot \alpha, X_{n+1}^2\cdot \alpha)       
      &\text{ += } R_1(\omega_{R_1}), \ldots, R_m(\omega_{R_m}) \label{eq:regression-tree} \\
     \text{where}\hspace{1em} &\alpha = \mathbf{1}_{X_1 \geq t_1} \cdot \mathbf{1}_{X_2 \leq t_3} \cdot
    \mathbf{1}_{X_5 \leq t_6} \nonumber
  \end{align}
\end{small}
The product aggregate $\alpha$ evaluates to 1 whenever all conditions in the subscript are satisfied and to 0 otherwise. 

For a categorical attribute $X$, the variance for all split
conditions can be expressed using a single query of the form~\eqref{eq:regression-tree} extended with the group-by attribute $X$.

For classification trees, the label $X_{n+1}$ has a set $\mbox{Dom}(X_{n+1})$ of categories. The cost is given by the entropy or Gini index:
\begin{small}
    \begin{align*}
      &\text{entropy} = - \hspace{-1.5em}\sum_{k\in \mbox{Dom}(X_{n+1})}
        \hspace{-1em}\pi_k \log (\pi_k)
      && \hspace{2em} \text{gini} =  1 - \hspace{-1.5em}\sum_{k\in \mbox{Dom}(X_{n+1})}
         \hspace{-1em}\pi_k^2
    \end{align*}
\end{small}
The aggregates $\pi_k$ for $k\in \mbox{Dom}(X_{n+1})=\{k_1,\ldots,k_p\}$ compute the frequencies of each category $k$ for the label $X_{n+1}$ in the dataset $D_i$, i.e., for category $k$ this frequency is the fraction of the tuples in $D_i$ where $X_{n+1}=k$ and of all tuples in $D_i$: $\frac{1}{|D_i|}\sum_{\mathbf{X} \in D_i} \mathbf{1}_{X_{n+1}=k}$.
These frequencies can all be computed with the following two aggregate queries:
\begin{small}
  \begin{align}
    CT(X_{n+1}; \alpha)&\text{ += } R_1(\omega_{R_1}), \ldots, R_m(\omega_{R_m})
                         \label{eq:classification-tree} \\
    CT(\alpha)&\text{ += } R_1(\omega_{R_1}), \ldots, R_m(\omega_{R_m})
                \label{eq:classification-tree2}
  \end{align}
\end{small}\vspace*{-1em}

For a categorical attribute $X$, the cost of all split
conditions can be expressed using two queries of the form~\eqref{eq:classification-tree} and~\eqref{eq:classification-tree2} extended with the group-by attribute $X$.


\subsubsection*{Applications need a large number of aggregates.} 
The number of aggregates in a batch is a function of the number $n$ of attributes in the database: $\frac{1}{2}(n+1)(n+2)$ for linear regression; $\frac{1}{2}\big[{\binom{n+d}{d}}^2 \!+\! {\binom{n+d}{d}}\big]$ for polynomial regression of degree $d$; 
$2^d\nu$ for $d$-dimensional data cubes with $\nu$ measures; $\frac{n(n-1)}{2}$ for Chow-Liu trees with $n$ nodes; and $dn(p+1)c$ for classification/regression trees with $d$ nodes where $c$ conditions are tried per attribute and the response has $p$ categories in case of classification tree; the formula for regression tree is obtained with $p=2$.
Table~\ref{table:numberaggregates} gives the number of aggregates for these applications and our four datasets, whose details are in Table~\ref{table:datasetstats}. This number ranges from tens to tens of thousands.

\subsubsection*{Further Applications}
Virtually any in-database machine learning setting can benefit from an efficient processor for aggregate batches over joins. Although not reported in this work, we also investigated SVM, k-means clustering, and low-rank models such as quadratically regularized PCA and Non-Negative Matrix Factorization, as well as linear algebra operations such as QR and SVD decompositions of matrices defined by the natural join of database relations. All these applications decompose into batches of aggregates of a similar form to those mentioned here.

\section{The LMFAO Engine}
\label{sec:engine}

In this section we discuss key design choices for LMFAO and motivate them using examples. Section~\ref{sec:introduction} already provided an overview of its layers that are depicted in Figure~\ref{fig:lmfao-architecture}.

\subsection{Aggregates over Join Trees}
\label{sec:jointrees}

LMFAO evaluates a batch of aggregate queries of the form~\eqref{eq:query} over a join tree of the database schema, or equivalently of the natural join of the database relations. We next recall the notion of join trees and exemplify the evaluation of aggregates over joins trees by decomposing them into views.

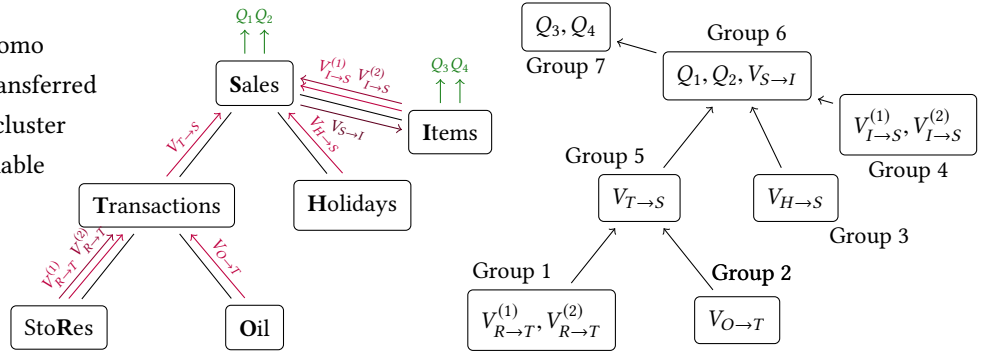
\begin{figure*}[t] 
  \subfigure{
    \begin{minipage}{1cm}
      \hspace*{-3.5em}
      \begin{align*}
        &\text{{\bf Sales:} \underline{date}, \underline{store}, \underline{item}, units, promo}\\
        &\text{{\bf Holidays:} \underline{date}, htype, locale, transferred}\\
        &\text{{\bf StoRes:} \underline{store}, city, state, stype, cluster}\\
        &\text{{\bf Items:} \underline{item}, family, class, perishable}\\
        &\text{{\bf Transactions:} \underline{date}, \underline{store}, txns}\\
        &\text{{\bf Oil:} \underline{date}, price}\\
        & \\
      \end{align*}
    \end{minipage}
  }
  \subfigure{
    \begin{minipage}{6cm}\hspace*{-1.5cm}
   \begin{tikzpicture}[yscale = 0.65, xscale=0.65, every node/.style={transform shape}]
     
     \tikzstyle{data} = [draw, rectangle, rounded corners = .07cm, align=center,
     inner sep = .2cm, outer sep = .1 cm]
    
    \node[data,scale=1.4] (sales) {\textbf{S}ales};
    \node[data,scale=1.4] at($(sales)+(-2,-2.5)$) (trans) {\textbf{T}ransactions};
    \node[data,scale=1.4] at($(trans)+(-2,-2.5)$) (store) {Sto\textbf{R}es};
    \node[data,scale=1.4] at($(trans)+(2,-2.5)$) (oil) {\textbf{O}il};        
    \node[data,scale=1.4] at($(sales)+(4,-1)$) (item) {\textbf{I}tems};
    \node[data,scale=1.4] at($(sales)+(2,-2.5)$) (holi) {\textbf{H}olidays};

    \draw (sales) -- (trans); 
    \draw (trans) -- (store); 
    \draw (trans) -- (oil);
    \draw (sales) -- (item);
    \draw (sales) -- (holi);

    \draw[->,color=purple] ($(trans.north)+(0.2, 0.01)$) -- ($(sales.south)+(-0.7, -0.01)$)
    node[midway,above,sloped,scale=1]{$V_{T\rightarrow S}$};
    
    \draw[->,color=purple] ($(store.north)+(0.2, 0.01)$) -- ($(trans.south)+(-0.7, -0.01)$);
    \draw[->,color=purple!80!black] ($(store.north)+(0, 0.01)$) -- ($(trans.south)+(-0.9, -0.01)$)
    node[midway,above,sloped,scale=1]{{\color{purple}$V_{R\rightarrow T}^{(1)}$} {\color{purple!80!black}$V_{R\rightarrow T}^{(2)}$}};

    \draw[->,color=purple] ($(oil.north)+(-0.2, 0.01)$) -- ($(trans.south)+(0.7, -0.01)$)
    node[midway,above,sloped,scale=1]{$V_{O\rightarrow T}$};

    \draw[->,color=purple] ($(holi.north)+(-0.2, 0.01)$) -- ($(sales.south)+(0.7, -0.01)$)
    node[midway,above,sloped,scale=1]{$V_{H\rightarrow S}$};

    \draw[->,color=purple] ($(item.north west)+(0, -0.15)$) -- ($(sales.east)+(0.01, -0.05)$);
    \draw[->,color=purple!80!black] ($(item.north west)+(0, 0)$) -- ($(sales.east)+(0.01, 0.1)$)
    node[midway,above,sloped,scale=1]{{\color{purple}$V_{I\rightarrow S}^{(1)}$} {\color{purple!80!black}$V_{I\rightarrow S}^{(2)}$}};
    
    \draw[->,color=purple!50!black] ($(sales.south east)+(0.01, 0.15)$) -- ($(item.west)+(0, 0.05)$)
    node[midway,below,sloped,scale=1]{$V_{S\rightarrow I}$};

    \draw[->,color=goodgreen] ($(sales.north)+(-0.2, 0.01)$) -- ($(sales.north)+(-0.2, 0.5)$)
    node[anchor=south]{$Q_1$};
    \draw[->,color=goodgreen] ($(sales.north)+(0.2, 0.01)$) -- ($(sales.north)+(0.2, 0.5)$)
    node[anchor=south]{$Q_2$};

    \draw[->,color=goodgreen] ($(item.north)+(-0.2, 0.01)$) -- ($(item.north)+(-0.2, 0.5)$)
    node[anchor=south]{$Q_3$};

    \draw[->,color=goodgreen] ($(item.north)+(0.2, 0.01)$) -- ($(item.north)+(0.2, 0.5)$)
    node[anchor=south]{$Q_4$};
  \end{tikzpicture}
  \end{minipage}
}
\subfigure{
\begin{minipage}{5cm}\hspace*{-1.7cm}
  \begin{tikzpicture}[yscale = 0.65, xscale=0.65, every node/.style={transform shape}]
     \tikzstyle{data} = [draw, rectangle, rounded corners = .07cm, align=center,
     inner sep = .2cm, outer sep = .1 cm]
    
    \node[data,scale=1.4] (sales) {$Q_1, Q_2, V_{S\rightarrow I}$};
    \node[data,scale=1.4] at($(sales)+(-2,-2.5)$) (trans) {$V_{T\rightarrow S}$};
    \node[data,scale=1.4] at($(trans)+(-2,-2.5)$) (store) {$V^{(1)}_{R\rightarrow T},V^{(2)}_{R\rightarrow T}$};
    \node[data,scale=1.4] at($(trans)+(2,-2.5)$) (oil) {$V_{O\rightarrow T}$};        
    \node[data,scale=1.4] at($(sales)+(3.5,-1)$) (item) {$V^{(1)}_{I\rightarrow S},V^{(2)}_{I\rightarrow S}$};
    \node[data,scale=1.4] at($(sales)+(1.2,-2.5)$) (holi) {$V_{H\rightarrow S}$};
    \node[data,scale=1.4] at($(sales)+(-3.5,1)$) (g7) {$Q_3, Q_4$};

    \draw[<-] (sales) -- (trans); 
    \draw[<-] (trans) -- (store); 
    \draw[<-] (trans) -- (oil); 
    \draw[<-] (sales) -- (item);
    \draw[<-] (sales) -- (holi);
    \draw[->] (sales) -- (g7);

    \node[anchor=south] at ($(sales.north)+(0.2, 0.5)$) {};

    \node[scale=1.4]  at ($(sales.north)+(0.2,0.2)$) (a) {Group 6};
    \node[scale=1.4]  at ($(trans.north)+(-0.7,0.2)$) (a) {Group 5};
    \node[scale=1.4]  at ($(store.north)+(-0.6,0.2)$) (a) {Group 1};
    \node[scale=1.4]  at ($(oil.north)+(0.3,0.3)$) (a) {Group 2};
    \node[scale=1.4]  at ($(item.south)+(0,-0.2)$) (a) {Group 4};
    \node[scale=1.4]  at ($(oil.north)+(0.3,0.3)$) (a) {Group 2};
    \node[scale=1.4]  at ($(holi.south east)+(0.5,-0.2)$) (a) {Group 3};
    \node[scale=1.4]  at ($(g7.south)+(0,-0.2)$) (a) {Group 7};

  \end{tikzpicture}
\end{minipage}
}

    \caption{(left) The schema for the Favorita dataset. (middle) A join tree for this schema with directional views and four queries, partitioned in 7 groups. (right) The dependency graph of the view groups.}
    \label{fig:join-tree-favorita}
  \end{figure*}

The {\em join tree} of the natural join of the database relations $R_1(\omega_{R_1}),\ldots,R_m(\omega_{R_m})$
is an undirected tree $T$ such that~\cite{abiteboul1995foundations}:
\begin{itemize}
\item The set of nodes of $T$ is $\{R_1,\ldots,R_m\}$.
\item For every pair of nodes $R_i$ and $R_j$, their common attributes are in the schema of every node $R_k$ along the distinct path from $R_i$ to $R_j$, i.e., $\omega_{R_i}\cap\omega_{R_j}\subseteq\omega_{R_k}$. 
\end{itemize}
Figure~\ref{fig:join-tree-favorita} shows a possible join tree for the natural join of the six relations in the Favorita dataset~\cite{favorita} (details are given in Appendix~\ref{appendix:datasets}). Instead of showing join attributes on the edges, we underline them in the schema (left) to avoid clutter.

Acyclic joins always admit join trees. Arbitrary joins are transformed into acyclic ones by means of hypertree decompositions and materialization of their nodes (called bags) using worst-case optimal join algorithms~\cite{Marx:2010,LFTJ}. We next exemplify the computation of aggregates over a join tree~\cite{BKOZ13,faq}.

\begin{example}\label{ex:pushdown1}
Let us compute the sum of the product of two aggregate functions $f(\text{units})$ and $g(\text{price})$
over the natural join of the Favorita relations:
\begin{small}
\begin{align*}
Q_1&(f(\text{units})\cdot g(\text{price})) \text{+=} 
S(\omega_S),T(\omega_T),R(\omega_R),O(\omega_O),H(\omega_H),I(\omega_I)
\end{align*}
\end{small}
We abbreviated the names of the Favorita relations as highlighted in Figure~\ref{fig:join-tree-favorita}. The aggregate functions $f$ and $g$ are over the attributes units in Sales and price in Oil.
We can rewrite $Q_1$ to push these functions down to the relations and also follow the structure of the join tree in Figure~\ref{fig:join-tree-favorita}:
\begin{small}
\begin{align*}
V_O(\text{date}; g(\text{price})) &\text{ += } O(\text{date}, \text{price})\\
V_R\text{(store}; 1)  &\text{ += } R(\text{store}, \text{city}, \text{state}, \text{stype}, \text{cluster})\\
V_T(\text{date},\text{store}; c\cdot p) &\text{ += } T(\text{date}, \text{store}, \text{t}), V_R\text{(store}; c),V_O(\text{date}; p)\\
V_H(\text{date}; 1) &\text{ += } H(\text{date}, \text{htype}, \text{locale}, \text{transferred})\\
V_I(\text{item}; 1) &\text{ += } I(\text{item}, \text{family}, \text{class}, \text{perishable})\\
Q_1(f(\text{units})\cdot p\cdot c_1\cdot c_2)  &\text{ += } V_T(\text{date},\text{store}; p), V_I(\text{item}; c_2), \\
&\hspace*{1.5em}V_H(\text{date}; c_1), S(\text{date}, \text{store}, \text{item}, \text{units}) 
\end{align*}
\end{small}
Except for $S$ and $O$, which have attributes in the aggregate functions, we only need to count the number of tuples with the same join key in each of the other relations.\qed
\end{example}

The computation of several aggregates over the same join tree may share views between themselves. 

\begin{example}\label{ex:pushdown2}
Consider now $Q_2(\text{family};g(\text{price}))$ over the same join. This query reports the sum of $g(\text{price})$ for each item family. We can rewrite it similarly to $Q_1$ in Example~\ref{ex:pushdown1}:
\begin{small}
\begin{align*}
V'_I(\text{family},\text{item}; 1) &\text{ += } I(\text{item}, \text{family}, \text{class}, \text{perishable})\\
Q_2(\text{family}; p\cdot c_1\cdot c_2)  &\text{ += } V_T(\text{date},\text{store}; p), V'_I(\text{family},\text{item}; c_2), \\
&\hspace*{1.5em}V_H(\text{date}; c_1),S(\text{date}, \text{store}, \text{item}, \text{units}) 
\end{align*}
\end{small}
We can share the views $V_T$, and thus its underlying views $V_O$ and $V_R$, and
$V_H$ between $Q_1$ and $Q_2$. \qed
\end{example}

\subsection{Directional Views over Join Trees}

An alternative evaluation for $Q_2$ in Example~\ref{ex:pushdown2} would not create the view $V'_I$ and instead create a view $V_S(\text{item}; p)$ over the subtree rooted at Sales and then join it with Items in the body of $Q_2$. This effectively means that we use the same join tree but rooted at different nodes: Sales for $Q_1$ and Items for $Q_2$. 
This also means that the edge between Sales and Item has two views, yet with different direction.

To accommodate this evaluation approach, we introduce {\em directional views}: These are queries of the form~\eqref{eq:query} where we also specify their direction. They flow along an edge from a source node to a neighboring target node and are computed using the relation at the source node and some of its incoming views. Examples~\ref{ex:pushdown1} and \ref{ex:pushdown2} showed views whose directions are always towards the root Sales. The direction of $V_I$ is $I\rightarrow S$ and of $V_S$ for the alternative evaluation of $Q_2$ is $S\rightarrow I$.

Consider a join tree $T$ with root $S$ and children $C_1,\ldots,C_k$, where child $C_i$ is the root of a subtree $T_i$ in $T$. We use $\omega_{C_i}$ and $\omega_{T_i}$ to denote the schema of the relation $C_i$ in $T$ and respectively the union of the schemas of all relations in $T_i$. 

We decompose a query $Q(F;\alpha)$ with group-by attributes $F$ and aggregate function $\alpha$ as follows:
\begin{small}
\begin{align*}
Q(F;\alpha) \text{ += } S(\omega_S),V_{C_1\rightarrow S}(F_1;\alpha_1),\ldots,V_{C_k\rightarrow S}(F_k;\alpha_k)
\end{align*}
\end{small}
The view $V_{C_i\rightarrow S}(F_i;\alpha_i)$ for a child $C_i$ of $S$ is defined as the ``projection'' of $Q$ onto $T_i$ as follows. Its group-by attributes are $F_i=(F\cap \omega_{T_i})\cup(\omega_S\cap\omega_{C_i})$; here, $\omega_S\cap\omega_{C_i}$ are the attributes shared between $S$ and a child $C_i$ and $F\cap\omega_{T_i}$ are the group-by attributes from $F$ present in $T_i$.  Its body is the natural join of the relations in $T_i$. If all attributes of $\alpha$ are (are not) in $\omega_{T_i}$, then $\alpha_i=\alpha$ (respectively $\alpha_i=1$). Otherwise, $T_i$ has some of the attributes required to compute $\alpha$, in which case we add them as group-by attributes, i.e., $F_i := F_i\cup (\omega_{T_i}\cap\omega_{\alpha})$, and use the aggregate $\alpha_i=1$ to count. We can now decompose the views $V_{C_i\rightarrow S}$ recursively as explained for $Q$.

Using different roots for different queries may lower the overall complexity of evaluating a batch of aggregates. At the same time, we would like to share computation as much as possible, which is intuitively maximized if all queries are computed at the same root. We next discuss our solution to this tension between complexity and sharing.

\subsection{Each Aggregate to Its Own Root}
\label{sec:root}

\nop{
\begin{itemize}
\item Discuss the importance to select roots for each query. 
\item We need a simple statement that shows the complexity advantage. 
\item We should show how we select roots for a query, show the greedy algorithm (?)
\end{itemize}
}

We next exemplify the advantage of evaluating a batch of queries, which are common in linear regression and mutual information settings where all attributes are categorical, at different roots in the join tree and then explain how to find a root for a given aggregate in a batch of aggregates.

\begin{example}
Consider the following count queries over the join of relations $S_k(X_k,X_{k+1})$ of size $N$, $\forall k\in[n-1]$:
\begin{small}
\begin{align*}
\forall i\in[n]: Q_{i} (X_i; 1) \text{ += } S_1(X_1,X_2),\ldots,S_{n-1}(X_{n-1},X_n).
\end{align*}
\end{small}
We first explain how to compute these $n$ queries  by decomposing them into directional views that are over the join tree $S_1-S_2-\cdots-S_{n-1}$ with root $S_1$ and have the same direction along this path towards the root.

For simplicity, we denote by $L^{i}_k$ the view constructed for $Q_{i}$ with direction from $S_k$ to $S_{k-1}$. The views are defined as follows, with the addition of $L^{n}_{n}(X_{n},X_{n}; 1) \text{ += } \text{Dom}(X_{n})$ that associates each value in the domain of $X_{n}$ with $1$.
\begin{small}
\begin{align*}
\forall k\in [i-1]:\ &L^{i}_{k}(X_k,X_i; c) &&\text{ += } S_k(X_k,X_{k+1}), L^{i}_{k+1}(X_{k+1},X_i; c)\\
\forall i\in[n-1]:\ &L^{i}_{i}(X_i,X_i; c) &&\text{ += } L^{i+1}_{i+1}(X_i,X_{i+1};c)\\
\forall i\in[n-1]:\ &Q_{i}(X_i; c) &&\text{ += } L^{i}_{1}(X_1,X_i;c)
\end{align*}
\end{small}
The above decomposition proceeds as follows. $Q_{n}$ counts the number of occurrences of each value for $X_n$ in the join. We start with 1, as provided by $L^{n}_n$, and progress to the left towards the root $S_1$. The view $L^{n}_{n-1}$ computes the counts for $X_n$ in the context of each value for $X_{n-1}$ as obtained from $S_{n-1}$. We need to keep the values for $X_{n-1}$ to connect with $S_{n-2}$. Eventually, we reach $L^n_1$ that gives the counts for $X_n$ in the context of $X_1$, and we sum them over the values of $X_1$. The same idea applies to any $Q_{i}$ with one additional optimization: Instead of starting from the leaf $S_{n-1}$, we can jump-start at $S_{i-1}$ and reuse the computation of the counts for $X_{i+1}$ in the context of $X_i$ as provided by $L^{i+1}_{i+1}$. We need $O(n^2)$ many views and those of them that have group-by attributes from two different relations take $O(N^2)$ time.

We can lower this complexity to $O(N)$ by using different roots for different queries. We show the effect of using the root $S_i$ for query $Q_{i}$. For each query $Q_i$, we construct two directional views: view $R_i$ from $S_{i-1}$ to $S_i$ (i.e., from left to right) and view $L_i$ from $S_{i+1}$ to $S_i$ (i.e., from right to left). The counts for $X_i$ values are the products of the counts in the left view $L_i$ and the right view $R_i$:
\begin{small}
\begin{align*}
\forall 1\leq i < n: L_i(X_i;c) &\text{ += } S_i(X_i,X_{i+1}), L_{i+1}(X_{i+1}; c)\\
\forall 1< i \leq n: R_i(X_i;c) &\text{ += } S_{i-1}(X_{i-1},X_i), R_{i-1}(X_{i-1}; c)\\
\forall 1\leq i \leq n: Q_{i}(X_i;c_1\cdot c_2) &\text{ += } L_i(X_i; c_1), R_i(X_i; c_2)
\end{align*}
\end{small}
We also use two trivial views $R_1(X_1;1) \text{ += } \text{Dom}(X_1)$ and  $L_n(X_n;1) \text{ += } \text{Dom}(X_n)$. Note how the left view $L_i$ is expressed using the left view $L_{i+1}$ coming from the node $S_{i+1}$ below $S_i$. Similarly for the right views. Each of the $2n$ views takes linear time. Moreover, they share much more computation among them than the views $L^i_k$ used in the first scenario.

The second approach that chooses the root $S_i$ for query $Q_i$ can also be used for queries over all pairs of attributes:
\begin{small}
\begin{align*}
\forall i,j\in[n]: Q_{i,j} (X_i,X_j; 1) \text{ += } S_1(X_1,X_2),\ldots,S_{n-1}(X_{n-1},X_n).
\end{align*}
\end{small}
Each of these $n^2$ queries takes time $O(N)$ for $|i-j|\leq 1$ and $O(N^2)$ otherwise. 
At each node $S_i$, we compute a left view $L_{i,j}$, for any $i<j\leq n$, that counts the number of tuples for $(X_i,X_j)$ over the path $S_i-\cdots-S_{n-1}$. Then, the overall count $c$ in $Q_{i,j}(X_i,X_j; c)$ is computed as the product of the count for $X_i$ given by the right view $R_i$ and the count for $(X_i,X_j)$ given by the left view $L_{i,j}$ ($\forall 1\leq i<j < n$):
\begin{small}
\begin{align*}
L_{i,j}(X_i,X_j;c) &\text{ += } S_i(X_i, X_{i+1}), L_{i+1,j}(X_{i+1},X_j; c)\\
Q_{i,j}(X_i,X_j;c_1\cdot c_2) &\text{ += } L_{i,j}(X_i, X_j; c_1), R_i(X_i; c_2)
\end{align*}
\end{small}
The trivial views $\forall i\in[n]: L_{i,i}(X_i,X_i;1) \text{ += } \text{Dom}(X_i)$ assign a count of 1 to each value of $X_i$. \qed
\end{example}

LMFAO chooses the root in a join tree for each query in a batch using a simple and effective approximation for the problem of minimizing the overall size of the views used to compute the entire batch. For each query $Q$ in the batch, we assign a weight to each relation $R$ in the join tree that is the fraction of the number of group-by attributes of $Q$ in $R$; if $Q$ has no group-by attribute, then any relation is a possible root and we assign to each relation the same weight that is an equal fraction of the number of relations. At the end of this weight assignment phase, each relation will be assigned some weight. We assign roots in the reverse order of their weights. A relation with the largest weight is then assigned as root to all queries that considered it as possible root. We break ties by choosing a relation with the largest size. The rationale for this choice is threefold. The choice for the largest relation avoids the creation of possibly large views over it. If the root for $Q$ has no group-by attribute of $Q$, then we will create views carrying around values for these attributes, so larger views. A root with a large weight ensures that many views share the same direction towards it, so their computation may be shared and they may be merged or grouped (as explained in the next sections).

\subsection{Merging and Grouping Views}
\label{sec:merging-views}

The views generated for a batch of aggregates can be consolidated or {\em merged} if they have in common: (1) only the group-by attributes and direction, (2) also the body, and (3) also the aggregates. The common case (3), which is also the most restrictive one, has been seen in Example~\ref{ex:pushdown2}: The same view is created for several queries, in which case we only need to compute it once. Case (2) concerns views with the same group-by attributes and join but different aggregates. Such views are merged into a single one that keeps the same group-by attributes and join but merges the lists of aggregates. Case (1) is the most general form of merging supported by LMFAO and consolidates the views into a new view that is a join of these views on their (same) group-by attributes. The reason why this merging is sound is twofold. First, these views are over the same join, so they have the same set of tuples over their group-by attributes. Second, the aggregates are functionally determined by the group-by attributes.

\begin{example}
We continue Examples~\ref{ex:pushdown1} and \ref{ex:pushdown2} and add a third count query $Q_3(\text{family}; h(\text{txns}, \text{city}))$ over the same join body as $Q_1$ and $Q_2$. This is decomposed into the following views over the same join tree rooted at Sales:
\begin{small}
\begin{align*}
&V'_O(\text{date};1) \text{ += } O(\text{date},\text{price})\\
&V'_R(\text{store}, \text{city}; 1) \text{ += } R(\text{store}, \text{city}, \text{state}, \text{stype}, \text{cluster})\\
&V'_T(\text{date},\text{store};\ h(\text{txns}, \text{city})\cdot c_1\cdot c_2) \text{ += } V'_R(\text{store}, \text{city}; c_1), \\
&\hspace{4em} T(\text{date}, \text{store}, \text{txns}), V'_O(\text{date};c_2)\\
&Q_3(\text{family}; c_1\cdot c_2\cdot c_3) \text{ += } S(\text{date},\text{store},\text{item},\text{units}),V_H(\text{date}; c_2), \\ 
&\hspace*{4em}V'_T(\text{date},\text{store}; c_1), V'_I(\text{family}, \text{item};c_2)
\end{align*}
\end{small}
where $V_H$ is shared with $Q_1$ and $Q_2$; $V'_I$ is shared with $Q_2$.
\nop{
}

The view $V_O(\text{date};g(\text{price}))$ for $Q_1$ can be merged with $V'_O(\text{date};1)$ for $Q_3$ into a new view $W_O$ since they have the same group-by attributes and body:
\begin{small}
\begin{align*}
W_O(\text{date};g(\text{price}),1) &\text{ += } O(\text{date},\text{price})
\end{align*}
\end{small}
Both views $V_T$ for $Q_1$ and $V'_T$ for $Q_3$ are now defined over $W_O$
instead of the views $V_O$ and $V'_O$. Views $V_T$ and $V'_T$ have the same
group by attributes and direction, but different bodies (one joins over $V_R$ and
the other over $V'_R$). We can merge them in $W_T$ following Case (1): 
\begin{small}
\begin{align*}
W_T(\text{date},\text{store}; a_1, a_2)  \text{ += } &
V_T(\text{date},\text{store}; a_1),
V'_T(\text{date},\text{store}; a_2) \qed
\end{align*}
\end{small}
\end{example}

Besides merging, {\em grouping} is another way of clustering the views that can share computation: We form groups of views that go out of the same node, regardless of their group-by attributes and bodies. We group the views as follows. We compute a topological order of these views: If a view $V_1$ uses a view $V_2$ in its body, i.e., it depends directly on it, then $V_1$ appears after $V_2$ in this order. We then traverse this order and create a group with all views such that (1) no view in the group depends on another view, and (2) all views within the group go out of the same relation in the join tree. 

Figure~\ref{fig:join-tree-favorita}(center) shows a scenario with directional views and four queries along the edges of our Favorita join tree. Their grouping is shown in Figure~\ref{fig:join-tree-favorita}(right).

In the next section, we show how to compute all views within a group in one scan over their common relation.

\subsection{Multi-Output Optimization}
\label{sec:multi-output}

The view group is a computational unit in LMFAO. We introduce a new optimization
that constructs a plan that computes all views in a group in one scan
over their common input relation. Since this plan outputs the results for several
views, we call it {\em multi-output optimization}, or MOO for short.

One source of complexity in MOO is that the views in the group are defined over
different incoming views. While scanning the common relation, the multi-output plan
looks up into the incoming views to fetch aggregates needed for the computation
of the views in the group.  A second challenge is to update the aggregates of
each view in the group as soon as possible and with minimal number of
computation steps.

MOO has three steps: (1) Find an order of join attributes of the common
relation; Register (2) incoming and outgoing views and (3) aggregate functions
to attributes in the attribute order. We next present each of
these steps and exemplify them using the following group of three views with the
common relation Sales ($S$):
\begin{small}
\begin{align*}
  & Q_4(f(\text{units})\cdot \alpha_1\cdot \alpha_4\cdot \alpha_8) \text{ += } S(\text{item}, \text{date}, \text{store}, \text{units}, \text{promo}), \\
  & \hspace*{2.5em}  V_T(\text{date}, \text{store}; \alpha_8), V_H(\text{date}; \alpha_4), V_I(\text{item};\alpha_1)  \\
  & Q_5(\text{store}; g(\text{item})\cdot h(\text{date}, \text{family})\cdot \alpha_4\cdot \alpha_8 \cdot \alpha_{13}) \text{ += } V_H(\text{date}; \alpha_4), \\
  & \hspace*{2.5em} V_T(\text{date}, \text{store}; \alpha_8), V_I^{'}(\text{item}, \text{family};\alpha_{13}), \\
  & \hspace*{2.5em}S(\text{item}, \text{date}, \text{store}, \text{units}, \text{promo}) \\
  & Q_6(\text{item}; g(\text{item})\cdot f(\text{units})\cdot \alpha_1 \cdot \alpha_4 \cdot  \alpha_8) \text{ += } V_T(\text{date}, \text{store}; \alpha_8), \\
  & \hspace*{2.5em}S(\text{item}, \text{date}, \text{store}, \text{units}, \text{promo}), 
    V_H(\text{date}; \alpha_4), V_I(\text{item};\alpha_1)
\end{align*}
\end{small}
{\bf\noindent Join attribute order.} The scan uses a total order on the join attributes of the relation $S$ and sees $S$ logically as a (partial) trie, grouped by the first join attribute and so on until the last join attribute. The leaves of the tries are relations over the remaining non-join attributes. This order can be computed offline. To avoid exploring all possible permutations of the join attributes, we proceed with the following approximation. We first compute the domain size for each join attribute in $S$, i.e., the number of its distinct values. We choose the order that is the increasing order in the domain sizes of these attributes: $\text{item}-\text{date}-\text{store}$. We sort $S$ in this order.

The multi-output execution plan uses a multi-way nested-loops join over the
relation and the incoming views, with one loop per join attribute. It sees the
incoming and outgoing views as functions that, for a given tuple over the
group-by attributes, look up the corresponding aggregate value. The aggregates
to compute are also functions, in particular sums of products of functions that
are UDAFs or lookups into incoming views. For instance, the aggregate of $Q_4$
is the product $f(\text{units})\cdot \alpha_1 \cdot \alpha_4 \cdot \alpha_8$,
where the last three components are provided by lookups in incoming views:
$\alpha_1 = V_I(i)$ for the aggregate $\alpha_1$ in the view $V_I$, where the
group-by attribute item is set to $i$; similarly for $\alpha_4 = V_H(d)$ and
$\alpha_8 = V_T(d,s)$.

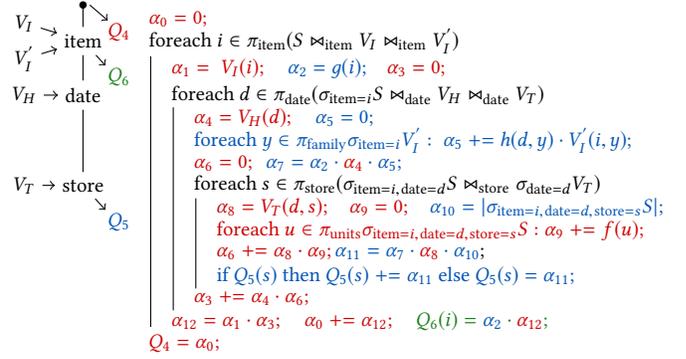
\begin{figure}
  \hspace*{-1em}%
  \begin{tikzpicture}[yscale=0.6, xscale=0.6, every node/.style={transform shape}]
    \node[scale=1.3,color=black,anchor=west] (int_item)
    {foreach $i \in \pi_{\text{item}}(S\Join_{\text{item}} V_{I}\Join_{\text{item}} V^{'}_{I})$};

    \node[scale=1.3,color=dred,anchor=west] at ($(int_item.west)+(0,0.5)$) (agg_item)
    {$\alpha_0 = 0;$};
    \node[scale=1.3,color=dred,anchor=west] at ($(int_item.west)+(0.5,-0.6)$) (agg_item)
    {$\color{dred}\alpha_1 =\; V_{I}(i); \hspace{1em} {\color{oxfordblue}\alpha_2 = g(i);} \hspace{1em} \alpha_3 = 0;$};
    
    \node[scale=1.3,color=black,anchor=west] at ($(int_item.west)+(.5,-1.2)$) (int_date)
    {foreach $d \in \pi_{\text{date}}( \sigma_{\text{item}=i}S \Join_\text{date} V_H \Join_\text{date} V_{T})$};

    \node[scale=1.3,color=dred,anchor=west] at ($(int_date.west)+(0.5,-0.5)$) (agg_date)
    {$\alpha_4 = V_{H}(d); \hspace{1em} {\color{oxfordblue}\alpha_5 = 0;}$};

    \node[scale=1.3,color=oxfordblue,anchor=west] at ($(int_date.west)+(0.5,-1)$) (agg_date)
    {$\text{foreach }y \in  \pi_{\text{family}}\sigma_{\text{item}=i}V^{'}_{I} : \;
      \alpha_5 \pluseq h(d,y) \cdot V^{'}_{I}(i,y);$};

    \node[scale=1.3,color=dred,anchor=west] at ($(int_date.west)+(0.5,-1.5)$)
    (agg_date)
    {$\alpha_6 = 0; \hspace{0.5em} {\color{oxfordblue}\alpha_7 = \alpha_2
        \cdot {\color{dred}\alpha_4} \cdot \alpha_5 ;}$};

    \node[scale=1.3,color=black,anchor=west] at ($(int_date.west)+(.5,-2)$) (int_store)
    {foreach $s \in \pi_{\text{store}}(\sigma_{\text{item}=i,\text{date}=d} S\Join_\text{store} \sigma_{\text{date}=d}V_T)$};

    \node[scale=1.3,color=dred,anchor=west] at ($(int_store.west)+(0.5,-0.5)$) (agg_store)
    {$\alpha_8 = V_{T}(d,s);\hspace{0.9em} \alpha_9 = 0; \hspace{0.9em} {\color{oxfordblue}\alpha_{10} = |\sigma_{\text{item}=i,\text{date}=d,\text{store}=s}S|;}$};

    \node[scale=1.3,color=dred,anchor=west] at ($(int_store.west)+(0.5,-1)$) (int_promo)
    {$\text{foreach }u \in \pi_\text{units}\sigma_{\text{item}=i,\text{date}=d,\text{store}=s} S: {\color{dred}\alpha_9 \pluseq f(u);}$};

    \node[scale=1.3,color=dred,anchor=west] at ($(int_store.west)+(0.5,-1.5)$)
    (agg_store) {$\alpha_6 \pluseq  \alpha_8 \cdot \alpha_9; {\color{oxfordblue}
        \alpha_{11} = \alpha_7 \cdot {\color{dred}\alpha_8} \cdot \alpha_{10}} ;$};
    
    \node[scale=1.3,color=oxfordblue,anchor=west] at
    ($(int_store.west)+(0.5,-2)$) (q2_out)
    {${\color{oxfordblue}\text{if } Q_5(s) \text{ then } Q_5(s) \pluseq \alpha_{11}
        \text{ else } Q_5(s) = \alpha_{11};}$};
    
    \node[scale=1.3,color=dred,anchor=west] at ($(int_store.west)+(0,-2.5)$) (rs1)
    {$\alpha_3 \pluseq \alpha_4 \cdot \alpha_6;$};
    
    \node[scale=1.3,color=dred,anchor=west] at ($(rs1.west)+(-0.5,-0.5)$) (rs2)
    {$\alpha_{12} = \alpha_1 \cdot \alpha_3; \hspace{1em} \alpha_0 \pluseq \alpha_{12};
      \hspace{1em} {\color{goodgreen} Q_6(i) =
        {\color{oxfordblue}\alpha_2} \cdot {\color{dred} \alpha_{12}};}$};
    \node[scale=1.3,color=dred,anchor=west] at ($(rs2.west)+(-0.5,-0.5)$) (rs3)
    {$Q_4 = \alpha_0;$};

    \node[scale=1.3,color=black] at($(int_item.west)+(-1.3,0)$)(item) {item};
    \node[scale=1.3,color=black] at($(item)+(0,-1.2)$) (date) {date};
    \node[scale=1.3,color=black] at($(date)+(0,-2)$) (store) {store};

    \draw   ($(item)+(0,0.8)$) -- (item) -- (date) -- (store); 
    \filldraw ($(item)+(0,0.8)$) circle (2pt);
    \node[scale=1.3,color=dred] at($(item)+(0,1)+(0.8,-0.8)$) (out_q4) {$Q_4$};

    \node[scale=1.3] at($(item)+(-1.3,0.4)$) (inc_item1) {$V_{I}$};
    \node[scale=1.3] at($(item)+(-1.3,-0.4)$) (inc_item2) {$V^{'}_{I}$};
    \node[scale=1.3,color=goodgreen] at($(item)+(0.8,-0.8)$) (out_item)
    {$Q_6$};

    \node[scale=1.3] at($(date)+(-1.3,0)$) (inc_holi) {$V_{H}$};       
    \node[scale=1.3] at($(store)+(-1.3,0)$) (inc_trans) {$V_{T}$};
    \node[scale=1.3,color=oxfordblue] at($(store)+(0.8,-0.8)$) (out_store) {$Q_5$};

    \draw[->] (inc_item1) -- (item);
    \draw[->] (inc_item2) -- (item);
    \draw[<-] ($(out_item.north west)+(0.1,-0.1)$) -- (item);

    \draw[->] (inc_holi) -- (date);
    
    \draw[->] (inc_trans) -- (store);
    \draw[<-] ($(out_store.north west)+(0.1,-0.1)$) -- (store);

    \draw[<-] ($(out_q4.north west)+(0.15,-0.1)$) -- ($(item)+(0.15,0.8)$);

    \draw[color=black] ($(int_item.south west)+(0.2,0.1)$)--
    ($(int_item.south west) + (0.2,-5.9)$);
    \draw[color=black] ($(int_date.south west)+(0.2,0.1)$)--
    ($(int_date.south west) + (0.2,-4.4)$);      
    \draw[color=black] ($(int_store.south west)+(0.2,0.1)$)--
    ($(int_store.south west) + (0.2,-1.9)$);

  \end{tikzpicture}
  \vspace*{-2em}
  \caption{Multi-output execution plan to compute {\color{dred}$Q_4$},
    {\color{oxfordblue}$Q_5$}, and {\color{goodgreen}$Q_6$} in the
    example of Section~\ref{sec:multi-output}.}
  \label{fig:mooexample}\vspace*{-1em}
\end{figure}

{\noindent\bf View registration.}  Each (incoming or outgoing) view $V$
is registered at the lowest attribute in the order that is a group-by attribute
of $V$. The reason is that at this attribute, all of the join
attributes that are group-by attributes of $V$ are fixed to constants and
we can construct the tuples over its group-by
attributes. The outgoing views without group-by attributes are registered outside the join attributes, as they are computed outside the outermost loop. Figure~\ref{fig:mooexample} depicts the registration of views in our example (left).


{\noindent\bf Aggregate function registration.}  Let $d_Q$ be the depth in the
attribute order where we registered an outgoing view $Q$.  We discuss the registration of
a product $p$ of aggregate functions in $Q$.  We decompose $p$ into minimal
partial products, such that no pair of functions from different partial products
depend on each other. Two functions depend on each other if they have non-join
attributes in the same relation or view:  In $Q_5$,
$h(\text{date},\text{family})$ and $V^{'}_I (\text{item},\text{family})$
depend on each other, because they share the non-join attribute $\text{family}$.
Dependent functions need to be evaluated together in loops over the distinct
values of the non-join attributes in the context of the values of the join
attributes. The reason for non-join attributes is that the
join attributes are fixed by the nested-loops join. The evaluation for each
partial product is to be performed at the attribute of largest depth that is a
parameter of any of the dependent functions.

If several functions in $p$ are registered at the same depth $d$, we multiply
their values. This is the partial product $p_d$ of functions in $p$ that can be
computed at depth $d$.  In order to obtain the final product $p$, we combine the
partial products that were computed at each depth $d$ as follows. If $d < d_Q$,
we register at $d$ one intermediate aggregate $a_d$ that is the product of $p_d$
and the intermediate aggregate $a_{d-1}$ that is computed at depth $d-1$. This
intermediate aggregate is computed {\em before} we proceed to the next attribute
in the order.  If $d > d_Q$, we register at $d$ a running sum $r_d$ over the
product of $p_d$ and the running sum $r_{d+1}$. The running sum is computed {\em
  after} we return from the next attribute in the order.  If $d = d_Q$, the
product $p$ corresponds to the product of $p_d$, the intermediate aggregate
$a_{d-1}$, and the running sum $r_{d+1}$. The product $p$ is computed {\em
  after} we return from depth $d+1$ in the order, and then added to the tuple
over the group-by attributes of $Q$.

\begin{example}
  Figure~\ref{fig:mooexample} depicts the computation of $Q_4$, $Q_5$, and
  $Q_6$. We register the components of the aggregate in $Q_4$ depending on the
  group-by attributes of their respective views: $f(\text{units})$ at store
  since the non-join attribute units is accessible once the join attributes are
  fixed in $S$; $\alpha_8$ also at store; $\alpha_4$ at date; $\alpha_1$
  at item.  The function $f(\text{units})$ has a special treatment, since units
  is not a join attribute. Within the context in relation $S$ of an item $i$,
  date $d$, and store $s$, we iterate over the qualifying tuples in $S$ and
  accumulate in the local variable $\alpha_9$ the sum over all values $f(u)$ for
  each value $u$ for units. Once we computed locally the values for the
  component aggregates, we combine them with a minimal number of computation
  steps. We use local variables for running sums of multiplications of these
  values. As soon as the aggregates $f(\text{units})$ and $\alpha_8$ are
  computed within the context of an item $i$, date $d$, and store $s$, we add
  their multiplication to a local variable $\alpha_6$; this variable is
  initialized to 0 outside the loop over stores and its content is accumulated
  in $\alpha_3$ right after the same loop.  This accumulation is also used for
  the loops over dates and then items. Since $Q_4$ has no group-by attributes,
  its result is the scalar representing the aforementioned accumulation:
  $Q_4=\alpha_0$.

  $Q_5$ and $Q_6$ are treated similarly to $Q_4$, with the difference that they
  have group-by attributes. We insert tuples in $Q_5$ within the loop over
  stores and update the aggregate value for a given store if the same store
  occurs under different (item, date) pairs. $Q_6$ reuses the aggregates
  $\alpha_{12}$ computed for $Q_4$ and $\alpha_2$ computed for $Q_5$. The tuples
  for $Q_6$ are constructed in the order of the items enumerated in the
  outermost loop.\qed
\end{example}

{\noindent \bf Code Generation.} Instead of registering and interpreting the
views and their aggregates at runtime, we generate succinct and efficient C++
code for the shared computation of many aggregates in a view group. This code
follows the multi-output plan similar to that in Figure~\ref{fig:mooexample} and
features code specialization and optimization.  Here are examples of code
optimization already present in the code in Figure~\ref{fig:mooexample}.  The
local variable $\alpha_{10}$ stores the size of a fragment of $S$. Since $S$ is
an array and sorted by item, date, and store, this fragment is a contiguous
range whose size can be provided right away without having to enumerate over it.
We do not allocate local variables if there is no need, e.g., the view lookup in
$\alpha_5$.  The optimization also distinguishes the different requirements for
data structures representing the results of $Q_5$ and $Q_6$. Since we iterate
over distinct items and $Q_6$ has one tuple per item, we can store $Q_6$ as a
vector where each new item value is appended. In contrast, the plan may
encounter the same store under different (item, date) pairs and therefore stores
$Q_5$ as a hashmap to support efficient out-of-order updates. Further
optimizations are highlighted in Appendix~\ref{appendix:compilation}.

\begin{table}[t]
\centering
\begin{tabular}{|l|r|r|r|r|}\hline
                         & \multicolumn{1}{|c|}{ R }
                         & \multicolumn{1}{|c|}{ F }
                         & \multicolumn{1}{|c|}{ Y }   
                         & \multicolumn{1}{|c|}{ T }     \\\hline
  Tuples in Database     & 87M   & 125M  & 8.7M  & 30M   \\
  Size of  Database      & 1.5GB & 2.5GB & 0.2GB & 3.4GB \\\hline
  Tuples in Join Result  & 86M   & 127M  & 360M  & 28M   \\
  Size of  Join Result   & 18GB  & 7GB   & 40GB  & 9GB   \\\hline
  Relations              & 5     & 6     & 5     & 10    \\
  Attributes             & 43    & 18    & 37    & 85    \\
  Categorical Attributes & 5     & 15    & 11    & 26    \\\hline
\end{tabular}
\caption{Characteristics of used datasets: Retailer (R), Favorita (F), Yelp (Y), and TPC-DS (T).}
\label{table:datasetstats}\vspace*{-2em}
\end{table}

\begin{table*}[t]
\centering
\begin{tabular}{|l|r|r|r|r||r|r|r|r||r|r|r|r||r|r|r|r|}\cline{2-17}
  \multicolumn{1}{c}{  }
 & \multicolumn{4}{|c||}{ Retailer }
 & \multicolumn{4}{|c||}{ Favorita }
 & \multicolumn{4}{|c||}{ Yelp }  
 & \multicolumn{4}{|c|}{ TPC-DS } \\\hline
 & \multicolumn{1}{|c|}{A+I}
 & \multicolumn{1}{|c|}{V}
 & \multicolumn{1}{|c|}{G}
 & \multicolumn{1}{|c||}{Size}
 & \multicolumn{1}{|c|}{A+I}
 & \multicolumn{1}{|c|}{V}
 & \multicolumn{1}{|c|}{G}
 & \multicolumn{1}{|c||}{Size}
 & \multicolumn{1}{|c|}{A+I}
 & \multicolumn{1}{|c|}{V}
 & \multicolumn{1}{|c|}{G}
 & \multicolumn{1}{|c||}{Size}
 & \multicolumn{1}{|c|}{A+I}
 & \multicolumn{1}{|c|}{V}
 & \multicolumn{1}{|c|}{G}
 & \multicolumn{1}{|c|}{Size}     \\\hline\hline
  CM
 & 814 + 654  & 34  & 7  & 0.1
 & 140 + 46   & 125 & 9  & 0.9
 & 730 + 309  & 99  & 8  & 1795
 & 3061 + 590 & 286 & 14 & 577    \\
  \hline
  RT
 & 3141 + 16  & 19  & 9  & 0.1
 & 270 + 20   & 26  & 11 & 0.1
 & 1392 + 16  & 22  & 9  & 39
 & 4299 + 138 & 52  & 17 & 0.2    \\
  \hline
  MI
 & 56 + 22    & 78  & 8  & 76
 & 106 + 35   & 141 & 9  & 10
 & 172 + 64   & 236 & 9  & 1759
 & 301 + 95   & 396 & 15 & 55     \\
  \hline
  DC
 & 40 + 8     & 12  & 5  & 3944
 & 40 + 7     & 13  & 6  & 5463
 & 40 + 7     & 13  & 5  & 1876
 & 40 + 12    & 17  & 10 & 3794   \\
  \hline
\end{tabular}
\caption{Number of application aggregates (A), additional intermediate aggregates synthesised by LMFAO (I), views (V),
  and groups of views (G) for each dataset and aggregate batches: covar
  matrix (CM), regression tree node (RT), mutual information (MI), and data cube
  (DC). The size on disk of the application aggregates is given in MB.}
\label{table:numberaggregates}
\vspace*{-2em}
\end{table*}

\begin{table*}[t]
\centering
\begin{tabular}{|ll||r@{\hskip 1.8em}r|r@{\hskip 1.8em}r|r@{\hskip 1.8em}r|r@{\hskip 1.8em}r|}\hline
  \multicolumn{2}{|l||}{Aggregate batch}
 & \multicolumn{2}{|c|}{Retailer}
 & \multicolumn{2}{|c|}{Favorita}
 & \multicolumn{2}{|c|}{Yelp}
 & \multicolumn{2}{|c|}{TPC-DS}                                                                                            \\\hline\hline
  Count
 & LMFAO   & 0.80     & 1.00$\times$     & 0.97     & 1.00$\times$   & 0.68      & 1.00$\times$          & 5.01     & 1.00$\times$         \\
 & DBX     & 2.38     & 2.98$\times$     & 4.04     & 4.15$\times$   & 2.53      & 3.72$\times$          & 2.84     & 0.57$\times$         \\ 
 & MonetDB & 3.75     & 4.70$\times$     & 8.11     & 8.32$\times$   & 4.37      & 6.44$\times$          & 2.84     & 0.57$\times$         \\ \hline
  Covar Matrix
 & LMFAO   & 11.87    & 1.00$\times$     & 38.11    & 1.00$\times$   & 108.81    & 1.00$\times$          & 274.55   & 1.00$\times$         \\
 & DBX     & 2,647.36 & 223.10$\times$   & 773.46   & 20.30$\times$  & 2,971.88  & 27.31$\times$         & 9,454.31 & 34.44$\times$        \\
 & MonetDB & 3,081.02 & 259.64$\times$   & 1,354.47 & 35.54$\times$  & 5,840.18  & 53.67$\times$         & 9,234.01 & 33.63$\times$        \\\hline
  Regression 
 & LMFAO   & 1.80     & 1.00$\times$     & 3.49     & 1.00$\times$   & 8.83      & 1.00$\times$          & 105.66   & 1.00$\times$         \\
   Tree Node 
 & DBX     & 3,134.67 & 1,739.55$\times$ & 431.11   & 123.58$\times$ & 2,409.59  & 272.97$\times$        & 2,480.49 & 23.48$\times$        \\
 & MonetDB & 3,395.00 & 1,884.02$\times$ & 674.06   & 193.23$\times$ & 13,489.20 & 1,528.11$\times$      & 3,085.60 & 29.20$\times$        \\\hline
  Mutual
 & LMFAO   & 30.05    & 1.00$\times$     & 111.68   & 1.00$\times$   & 345.35    & 1.00$\times$          & 252.96   & 1.00$\times$         \\
  Information
 & DBX     & 178.03   & 5.92$\times$     & 596.01   & 5.34$\times$   & 794.00    & 2.30$\times$          & 1,002.84 & 3.96$\times$         \\
 & MonetDB & 297.30   & 9.89$\times$     & 1,088.31 & 9.74$\times$   & 1,952.02  & 5.65$\times$          & 1,032.17 & 4.08$\times$         \\\hline
  Data Cube
 & LMFAO   & 15.47    & 1.00$\times$     & 22.85    & 1.00$\times$   & 23.75     & 1.00$\times$  & 15.65    & 1.00$\times$ \\
 & DBX     & 100.08   & 6.47$\times$     & 273.10   & 11.95$\times$  & 156.67    & 6.60$\times$  & 66.12    & 4.23$\times$ \\
 & MonetDB & 111.08   & 7.18$\times$     & 561.03   & 24.55$\times$  & 260.39    & 10.96$\times$ & 74.38    & 4.75$\times$ \\\hline
\end{tabular}
\caption{Time performance (seconds) for computing various batches of aggregates
  using LMFAO, MonetDB, and DBX and the relative speedup of LMFAO over MonetDB and DBX.}
\label{table:aggbench}
\vspace*{-1.5em}
\end{table*}

 \section{Experiments}
\label{sec:experiments}

We conducted two types of performance benchmarks on four datasets: (1) the
computation of batches of aggregates in LMFAO, MonetDB, and DBX (a commercial
DBMS); and (2) the training of machine learning models in LMFAO, TensorFlow,
MADlib, and AC/DC.  The outcome of these experiments is twofold: (1) MonetDB and DBX cannot efficiently compute large batches of aggregates as
required by a variety of analytics workloads; (2) Scalability challenges faced
by state-of-the-art machine learning systems can be mitigated by a combination
of database systems techniques.

{\bf\noindent Datasets} We consider four datasets (Table~\ref{table:datasetstats}): (1) {\em
  Retailer}~\cite{ANNOS:DEEM:18} is used by a large US retailer for forecasting
user demands and sales; (2) {\em Favorita}~\cite{favorita} is a public real
dataset that is also used for retail forecasting; (3) {\em Yelp} is based on the
public Yelp Dataset Challenge~\cite{yelpdataset} and contains information about
review ratings that users give to businesses; (4) {\em TPC-DS}~\cite{tpcds}
(scale factor 10, excerpt) is a synthetic dataset designed for decision support
applications. The structure and size of these datasets are common in retail and
advertising, where data is generated by sales transactions or click
streams. Retailer and TPC-DS have a snowflake schema, Favorita has a star
schema. They have a large fact table in the center and several smaller dimension
tables. Yelp also has a star schema, but with many-to-many joins that
increase the size of the join result significantly compared to the input
database.

Appendix~\ref{appendix:datasets} gives a detailed description of each dataset,
including its schema and the join tree used for our experiments.
Appendix~\ref{appendix:experiments} details the experimental setup for each
workload and the limitations of the competing systems.

\subsection{Computing Batches of Aggregates}
\label{sec:aggcomp}

We compute the batches of aggregates for the following workloads and each of the
four datasets: (1) the covar matrix; (2) a single node in a regression tree; (3)
the mutual information of all pairwise combinations of discrete attributes; and
(4) a data cube. For each workload and dataset,
Table~\ref{table:numberaggregates} details how many aggregates, views, and
groups are computed. It also gives the size on disk of the aggregates. This is a
strong indicator of the running time; except for data cubes, these sizes are
much smaller than for the underlying join.

{\bf\noindent Competitors} We benchmarked our system LMFAO, its predecessor 
AC/DC \cite{ANNOS:DEEM:18}, MonetDB 1.1 \cite{monetdb},
and DBX (a commercial DBMS).  PostgreSQL (PSQL) 11.1 proved consistently slower than DBX and MonetDB. EmptyHeaded~\cite{emptyheaded} failed to compute our workloads (Appendix~\ref{appendix:experiments}).

{\bf\noindent Takeaways} Table~\ref{table:aggbench} presents the performance of
each system for the four workloads and for the count query, which is used to
assess the performance of many queries relative to this simple query. LMFAO
consistently outperforms both DBX and MonetDB on all experiments, with a speedup
of up to three orders of magnitude.  The reason is as follows. Whereas DBX and
MonetDB compute each individual query efficiently, they do not share computation
across them. In contrast, LMFAO clusters the query batch into a few groups
that are computed together in a single pass over the fact table and at most two
passes over the smaller dimension tables. The fact table in Retailer has few
attributes and most aggregates are computed over the dimension tables. In
comparison, Favorita requires relatively few aggregates, and the large fact 
table in TPC-DS has many attributes and thus more aggregates are computed over it.
This explains the relatively lower performance improvement for Favorita and
TPC-DS. For Yelp, LMFAO's decomposition of aggregates into views avoids the
materalization of the many-to-many joins.

\nop{
}

The performance gap is particularly large for regression tree nodes, which, in
contrast to the other workloads, do not require queries with
group-by attributes from different relations. LMFAO merges all aggregates into
few views and shares their computation over each input relation.

We use the count query to show how much computation is shared in LMFAO. For
instance, the covar matrix for Retailer has 814 aggregates.  Without sharing,
the performance would be at least $814\times$ that of the count query, or 6510
seconds. The performance of LMFAO is, however, $55\times$ better!

\nop{
}

\begin{figure}\hspace*{-1em}
  \includegraphics[width=.5\textwidth]{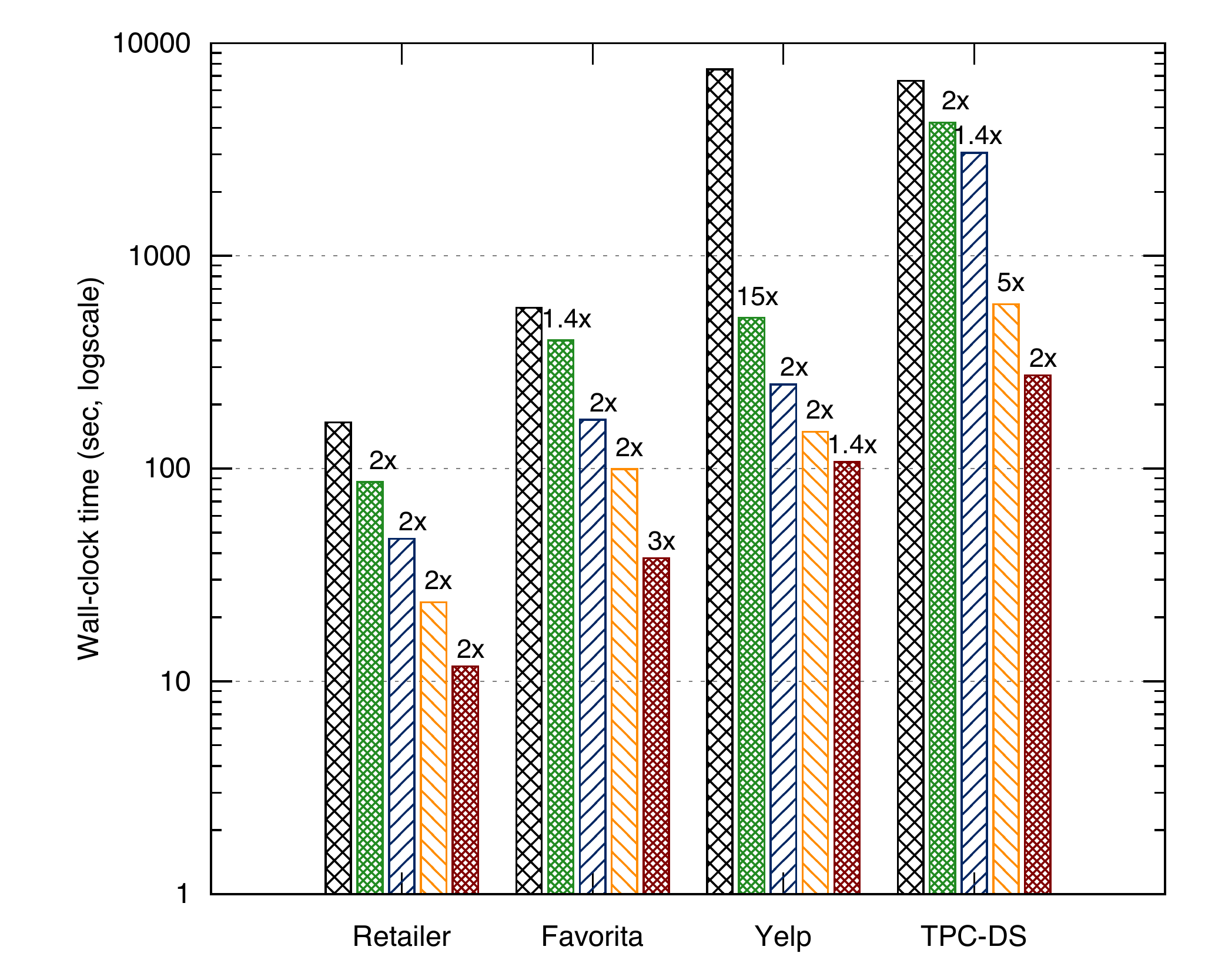}
  \vspace*{-2.5em}
  \caption{Performance impact of optimizations in LMFAO for computing the covar
    matrix. From left to right: no optimization (time for AC/DC proxy shown); compilation; plus multi-output; plus multi-root; and plus parallelization with 4 threads.
 Bars are
    annotated by relative speedup over the previous bar.}
  \nop{
  }
  \label{fig:lmfaobreakdown}
  \vspace*{-1em}
\end{figure}

{\bf\noindent LMFAO Optimizations}
Figure~\ref{fig:lmfaobreakdown} shows the performance benefit of
LMFAO optimizations for computing the covar matrix. The baseline is
its predecessor AC/DC (leftmost bar), a proxy for 
LMFAO without optimizations.  LMFAO with compilation but without the other optimizations achieves a speedup of $1.4 - 15\times$ over AC/DC. Multi-output and multiple roots together further
improve the performance by $4 - 7\times$ over LMFAO with compilation. Parallelization with four cores further improves the performance by 
$1.4 - 3\times$.

\nop{ Figure~\ref{fig:lmfaobreakdown} shows that the combination of several
  optimizations explains the superior performance of LMFAO for computing the
  covar matrix.  Most notably, the compilation improves the performance up to
  $15\times$ for Yelp. Multi-output and multiple roots together further improve
  the performance by roughly $4\times$ to $7\times$ for each
  dataset. Parallelization with four cores further improves the performance by
  up to $3\times$.  }

{\bf\noindent Compilation Overhead} The compilation overhead of LMFAO depends on
the workload. Using g++6.4.0 and eight cores, it ranges from 2
seconds for data cubes over Favorita to 50 seconds for the mutual information
batch over TPC-DS. This overhead is not reported in Table~\ref{table:aggbench} (we report the average of four subsequent runs). It can be reduced using LLVM code generation and
compilation~\cite{NeumannCompilation}.

\subsection{Training Models}
\label{sec:exp3}
  
We report the end-to-end performance of LMFAO for learning three machine
learning models: (1) ridge linear regression model; (2) regression tree; and
(3) classification tree. Models (1) and (2) are computed over Retailer and
Favorita, and used to predict the number of inventory units and respectively
number of units sold. Model (3) is learned over TPC-DS and used to predict
whether a customer is a preferred customer, as proposed in the Relational
Dataset Repository~\cite{ctu:repo}. To assess the accuracy of the models, we
separate out a test dataset. The training dataset for each model is defined by
the natural join of the remaining tuples in the input database.

{\bf\noindent Competitors} We benchmarked LMFAO against several analytics
tools commonly used in data science: TensorFlow 1.12
(compiled with AVX optimization enabled)~\cite{tensorflow}, MADlib
1.8~\cite{MADlib:2012}, R~\cite{R-project}, scikit-learn 0.20~\cite{scikit2011}, and Python StatsModels~\cite{P-StatsModels}. The latter
three fail to compute the models either due to internal design
limitations or out-of-memory error. TensorFlow mitigates this issue by using
an iterator interface that only loads a small batch of tuples at a time. MADlib
is an in-database analytics engine integrated with PSQL.

We also compared against AC/DC~\cite{ANNOS:DEEM:18} for learning linear
regression models over databases. Apart of AC/DC, all other systems require the
full materialization of the training dataset. In addition, TensorFlow requires a
random shuffling of the training dataset for linear regression models. We used
PSQL to compute the join and the shuffling steps.

{\bf\noindent Takeaways} Tables~\ref{table:regressionbench} and
\ref{table:classificationbench} give the performance of the systems. LMFAO is
able to compute all models orders-of-magnitude faster than the competitors.
For Retailer and Favorita, LMFAO learns the model over the input database even
faster than it takes PSQL to compute the join. This is because LMFAO avoids the
materialization of the large training dataset and works directly on the input
database: For Retailer, the former is $10\times$ larger than the latter
(Table~\ref{table:datasetstats}).  Furthermore, for linear regression, the
convergence step takes as input the result of the aggregate batch, which is
again at least an order of magnitude smaller than the input database.
    
LMFAO learns the linear regression models with the same accuracy as the
closed-form solution computed by MADlib yet in a fraction of the time it takes
MADlib. Tensorflow takes orders of magnitude longer for one epoch (one pass over
the training dataset).  The model that Tensorflow learns for Favorita also has
comparable accuracy to the closed-form solution, but for Retailer the
root-mean-square-error of the model is only marginally better than a baseline
model which always predicts the average of the label over the training
dataset\footnote{The error analysis for TensorFlow was updated after the
  original paper was published.}. TensorFlow would require more epochs to
converge to the solution of LMFAO. LMFAO also outperforms the specialized AC/DC
engine by up to $18\times$.

LMFAO learns decision trees with the same accuracy orders-of-magnitude faster
than MADlib. TensorFlow times out after 12 hours in all cases; we show the time
to compute the tree root as indication of the overall runtime.

\begin{table}[t]
\centering 
\begin{tabular}{|ll||r|r|}\hline
                      &                & Retailer  & Favorita  \\\hline\hline
  Join                & PSQL           & 152.06    & 129.32    \\
  Join Shuffle        & PSQL           & 5,488.73  & 1,720.02  \\
  Join Export         & PSQL           & 351.76    & 241.03    \\\hline
  \multicolumn{4}{|c|}{Linear Regression}                      \\\hline
  TensorFlow          & (1 epoch)      & 7,249.58  & 4,812.01  \\
  MADlib              &                & 5,423.05  & 19,445.58 \\
  AC/DC               &                & 110.88    & 364.17    \\
  LMFAO               &                & 6.08      & 21.23     \\\hline
  \multicolumn{4}{|c|}{Regression Trees}                       \\\hline
  TensorFlow          & (1 node)       & 7,773.80  & 20,368.73 \\
  MADlib              & (max 31 nodes) & 13,639.84 & 19,839.12 \\
  LMFAO               & (max 31 nodes)     & 21.28     & 37.48     \\\hline
\end{tabular}

\caption{Time performance (seconds) for learning linear regression models and
  regression trees over Favorita and Retailer.}
\label{table:regressionbench}
\vspace*{-2.5em}
\end{table}

\begin{table}[t]
  \centering 
  \begin{tabular}{|ll||r|}\hline
                &                & TPC-DS      \\\hline\hline
    Join        & PSQL           & 219.04      \\
    Join Export & PSQL           & 350.02      \\\hline
    \multicolumn{3}{|c|}{Classification Trees} \\\hline
    TensorFlow  & (1 node)       & 10,643.18  \\
    MADlib      & (max 31 nodes) & 34,717.63   \\
    LMFAO       & (max 31 nodes) & 720.86     \\\hline
  \end{tabular}
\caption{Time performance (seconds) for learning classification trees over TPC-DS.}
\label{table:classificationbench}
\vspace*{-2.5em}
\end{table}

\section{Related Work}
\label{sec:relatedwork}

LMFAO builds on a vast literature of database research. We cited highly relevant work in previous sections. We next mention further connections to work on sharing computation and data systems for learning models. LMFAO computes a batch of group-by aggregates over the same joins without materializing these joins, in the spirit of ad-hoc mining~\cite{Chaudhuri:DMDB:1998}, eager aggregation~\cite{Larson:LazyEager:1995}, and factorized databases~\cite{BKOZ13}.

{\bf\noindent Sharing Computation}
 Prior techniques for data cubes use a lattice of sub-queries to capture sharing across the group-by aggregates defining data cubes~\cite{DataCube:SIGMOD:1996,DataCube:SIGMOD:1997}. Which cells to materialize in a data cube is decided based on space or user-specified constraints~\cite{DataCube:SIGMOD:1996,DataCube:SIGMOD:1997}. \nop{In our setting, we need to always compute all aggregates needed for training machine learning models.} More recent work revisited shared workload optimization for queries with hash joins and shared scans and proposes an algorithm that, given a set of statements and their relative frequency in the workload, outputs a global plan over shared operators~\cite{ETH:Shared:2014}. Data Canopy is a library of frequently used statistics in the form of aggregates that can speed up repeating requests for the same statistics. It is concerned with how to decompose, represent, and access such statistics in an efficient manner~\cite{DataCanopy:SIGMOD:2017}.
 
{\bf\noindent Multi-Query Optimization (MQO)}~\cite{MQO:TODS:1988} is concerned with identifying common subexpressions across a set of queries with the purpose of avoiding redundant computation. One of the three types of view merging in LMFAO is also concerned with the same goal, though for directional views with group-by aggregates. LMFAO's view merging proved useful in case of very many and similar views, such as for the applications detailed in Section~\ref{sec:applications}. An alternative type of MQO is concerned with caching intermediate query results, such as in the MonetDB system that we used in experiments.

{\bf\noindent Learning over Multi-Relational Data}
There are {\em structure-agnostic} and {\em structure-aware} solutions depending on whether they  exploit the structure in the data or not.

The {\em structure-agnostic} solutions are by far the most common. They first construct the training dataset using a data system capable of computing queries and then learn the model over the materialized training dataset using 
an ML library or statistical package. The first step is performed in Python Pandas, R dplyr, or database systems such as PostgreSQL and SparkSQL~\cite{Spark:NSDI:2012}. The second step commonly uses  scikit-learn~\cite{scikit2011}, Python StatsModels~\cite{P-StatsModels}, TensorFlow~\cite{tensorflow}, R~\cite{R-project}, MLlib~\cite{MLlib:JMLR:2016}, SystemML~\cite{Boehm:2016}, or XGBoost~\cite{xgboost}. 
Although one could combine any data system and ML library, working
solutions feature combinations that avoid the expensive data export/import at
the interface between the two systems, e.g., MLlib over SparkSQL, the Python
packages over Pandas, R over dplyr, and 
MADlib~\cite{MADlib:2012} over PostgreSQL. MADlib,
Bismarck~\cite{Bismarck:SIGMOD:2012}, and GLADE PF-OLA~\cite{Rusu:2015}
define ML tasks as user-defined aggregate
functions (UDAFs). Although UDAFs share the same execution space with the
query computing the training dataset, they are treated as black boxes and executed after the training dataset is materialized.

A disadvantage of two-step solutions is the required materialization of the
training dataset that may be much larger than the input data (cf.\@
Table~\ref{table:datasetstats}). This is exacerbated by the stark asymmetry
between the two steps: Whereas data systems tend to scale to large datasets,
this is not the case for ML libraries. Yet, the two-step solutions expect by
design that the ML libraries work on even larger inputs than the data systems!
A further disadvantage is that these solutions inherit the limitations of both
underlying systems.  For instance, the R data frame can host at most $2^{31}$
values, which makes it impossible to learn models over large datasets, even if
data systems can process them. Database systems can only handle up to a few
thousand columns per relation, which is usually smaller than the number of
features of the model.

The {\em structure-aware} solutions tightly integrate the dataset construction and the learning steps, and allow the second step to exploit the relational structure in the input data. There is typically one unified execution plan for both the feature
extraction query and the subsequent learning task, with subcomponents of the
latter possibly pushed past the joins in the former. This plan works directly on the input data and computes sufficient statistics of much smaller size than of the training dataset (cf.\@ Table~\ref{table:numberaggregates}).  Our system LMFAO is a prime example of this class. It builds on F~\cite{SOC:SIGMOD:16} and AC/DC~\cite{ANNOS:DEEM:18}. F supports linear regression models. AC/DC generalizes F to non-linear models, categorical features, and model reparameterization under functional dependencies. A key aspect that sets apart F, AC/DC, and LMFAO from
all other efforts is the use of execution plans for the mixed workload of queries and learning whose complexity may be asymptotically lower even than that of the materialization step. In particular, this line of work shows that all machine learning approaches that require as input the materialization of the result of the feature extraction query may be asymptotically suboptimal.

Further examples in this category are: Orion~\cite{KuNaPa15} and
Hamlet~\cite{Kumar:SIGMOD:16}, which support generalized linear models and
Na\"ive Bayes classification; recent efforts on scaling linear algebra using
existing distributed database systems~\cite{LAoverDBMS:SIGREC:2018}; the
declarative language BUDS~\cite{BUDS:SIGMOD:2017}, whose compiler can perform
deep optimizations of the user's program; and Morpheus~\cite{Kumar:PVLDB:2017}.
Morpheus factorizes the computation of linear algebra operators summation,
matrix-multiplication, pseudo-inverse, and element-wise operations over training
datasets defined by key-foreign key star or chain joins. It represents the
training dataset as a normalized matrix, which is a triple of the fact table, a
list of dimension tables, and a list of indicator matrices that encode the join
between the fact table and each dimension table. Morpheus provides operator
rewritings that exploit the relational structure by pushing computation past
joins to the smaller dimension tables. Initial implementations of Morpheus are
built on top of the R and Python numpy linear algebra packages.  Morpheus only
supports key-foreign key star or chain joins and models that are expressible in
linear algebra.  In contrast, LMFAO supports arbitrary joins and rich aggregates
that can capture computation needed by a large heterogeneous set of models
beyond those expressible in linear algebra, including, e.g., decision trees.
\nop{
}

{\bf\noindent Optimizations in ML packages} Most ML libraries exploit sparsity in the form of zero-values (due to missing values or one-hot encoding), yet are not structure-aware. LMFAO exploits a more powerful form of sparsity that is prevalent in training datasets defined by joins of multiple relations: This is the join factorization that avoids the repeated representation of and computation over arbitrarily-sized data blocks. LMFAO's code optimizations aim specifically at generating succinct and efficient C++ code for the shared computation of many aggregates over the join of a large table and several views represented as ordered vectors or hashmaps. The layout of the generated code is important: how to decompose the aggregates, when to initialize and update them, how to share partial computation across many aggregates with different group-by and UDAFs (Section~\ref{sec:multi-output}). Lower-level optimizations (Appendix ~\ref{appendix:compilation}) are generic and adapted to our workload, e.g., how to manage large amounts of aggregates and how to update them in sequence.
LMFAO's multi-aggregate optimizations are absent in ML and linear algebra packages. We next highlight some code optimizations used in these packages. BLAS and LAPACK provide cache-efficient block matrix operations.
Eigen~\cite{eigen} supports both dense and sparse matrices, fuses operators to avoid intermediate results, and couples loop unrolling with SIMD vectorization. 
SPOOF~\cite{SPOOF:CIDR:2017} translates linear algebra operations into
sum-product form and detects opportunities for aggregate pushdown and operator fusion.
LGen~\cite{Spampinato:2016} uses compilation to generate efficient basic linear algebra operators for small dense, symmetric, or triangular matrices by employing loop fusion, loop tiling, and vectorization. TACO~\cite{kjolstad:2017:taco} can generate compound linear algebra operations on both dense and sparse matrices.
LMFAO can also learn decision trees, which cannot be expressed in linear algebra. XGBoost~\cite{xgboost} is a gradient boosting
library that uses decision trees as base learners. It represents the training
dataset in a compressed sparse columnar (CSC) format, which is partitioned into blocks that are optimized for cache
access, in-memory computation, parallelization, and can be stored on disk for out-of-core learning. LMFAO may also benefit from a combination of value-based compression and factorized representation of the training dataset, as well as from an out-of-core learning mechanism.


\begin{acks}
  This project has received funding from the European Union's Horizon 2020 research and innovation programme under grant agreement No 682588. 
 Olteanu  acknowledges a Google research award and an Infor research gift.  
 Nguyen gratefully acknowledges support from NSF grants CAREER DMS-1351362 
 and CNS-1409303, Adobe Research and Toyota Research, and a Margaret and 
 Herman Sokol Faculty Award. 
 Schleich acknowledges the AWS Proof of Concept Program.
\end{acks}

\bibliographystyle{ACM-Reference-Format}
\bibliography{bibtex}

\appendix

\begin{figure*}[t] \hspace*{-1em}
  \subfigure[Retailer]{
    \begin{tikzpicture}[yscale = 0.6, xscale=0.6, every node/.style={transform shape}]
       \tikzstyle{data} = [draw, rectangle, rounded corners = .07cm, align=center,
     inner sep = .2cm, outer sep = .1 cm]
    
      [yscale = 0.6, xscale=0.6, every node/.style={transform shape}]
      \node[data,scale=1.2] (inv) {Inventory};
      \node[data,scale=1.2] at($(inv)+(-2,-2)$) (loc) {Location};
      \node[data,scale=1.2] at($(loc)+(0,-2)$) (census) {Census};        
      \node[data,scale=1.2] at($(inv)+(3,-.5)$) (item) {Items};
      \node[data,scale=1.2] at($(inv)+(1,-2)$) (weat) {Weather};

      \draw (inv) -- (loc); 
      \draw (loc) -- (census); 
      \draw (inv) -- (item);
      \draw (inv) -- (weat);
    \end{tikzpicture}
  }\hfill
  \subfigure[Favorita]{
    \begin{tikzpicture}[yscale = 0.6, xscale=0.6, every node/.style={transform shape}]
       \tikzstyle{data} = [draw, rectangle, rounded corners = .07cm, align=center,
     inner sep = .2cm, outer sep = .1 cm]
    
      \node[data,scale=1.2] (sales) {Sales};
      \node[data,scale=1.2] at($(sales)+(-1.5,-2)$) (holi) {Holidays};
      \node[data,scale=1.2] at($(sales)+(1.5,-2)$) (trans) {Transactions};
      \node[data,scale=1.2] at($(sales)+(-1,2)$) (store) {Stores};
      \node[data,scale=1.2] at($(sales)+(1,2)$) (oil) {Oil};        
      \node[data,scale=1.2] at($(sales)+(2,0)$) (item) {Items};

      \draw (sales) -- (trans); 
      \draw (sales) -- (store); 
      \draw (sales) -- (oil); 
      \draw (sales) -- (item);
      \draw (sales) -- (holi);
    \end{tikzpicture}
  }\hfill
  \subfigure[Yelp]{
    \begin{tikzpicture}[yscale = 0.6, xscale=0.6, every node/.style={transform shape}]
       \tikzstyle{data} = [draw, rectangle, rounded corners = .07cm, align=center,
     inner sep = .2cm, outer sep = .1 cm]
    
      \node[data,scale=1.2] (sales) {Review};
      \node[data,scale=1.2] at($(sales)+(-1.5,-2)$) (trans) {Business};
      \node[data,scale=1.2] at($(sales)+(-1.5,2)$) (store) {Attribute};
      \node[data,scale=1.2] at($(sales)+(1.5,2)$) (oil) {Category};        
      \node[data,scale=1.2] at($(sales)+(1.5,-2)$) (holi) {User};

      \draw (sales) -- (trans); 
      \draw (sales) -- (store); 
      \draw (sales) -- (oil); 
      \draw (sales) -- (holi);
    \end{tikzpicture}
  }
  \hfill
  \subfigure[\label{fig:tpcds:schema}TPC-DS]{
    \begin{tikzpicture}[yscale = 0.6, xscale=0.6, every node/.style={transform shape}]
       \tikzstyle{data} = [draw, rectangle, rounded corners = .07cm, align=center,
     inner sep = .2cm, outer sep = .1 cm]
    
      \node[data,scale=1.2] (sales) {Store Sales};
      \node[data,scale=1.2] at($(sales)+(-2,-1.3)$) (cust) {Customer};
      \node[data,scale=1.2] at($(cust)+(-1.2,-1.3)$) (cdemo) {C\_Address};
      \node[data,scale=1.2] at($(cust)+(1.2,-1.3)$) (caddr) {C\_Demo};
      \node[data,scale=1.2] at($(sales)+(0,1.3)$) (date) {Date};
      \node[data,scale=1.2] at($(sales)+(1.75,1.3)$) (time) {Time};        
      \node[data,scale=1.2] at($(sales)+(-1.75,1.3)$) (item) {Item};
      \node[data,scale=1.2] at($(sales)+(-2.5,0)$) (store) {Store};
      \node[data,scale=1.2] at($(sales)+(2,-1.3)$) (hdemo) {H\_Demo};
      \node[data,scale=1.2] at($(hdemo)+(0,-1.3)$) (iband) {Inc\_Band};

      \draw (sales) -- (cust); 
      \draw (sales) -- (item); 
      \draw (sales) -- (date); 
      \draw (sales) -- (time);
      \draw (sales) -- (hdemo);
      \draw (sales) -- (store);
      \draw (cust) -- (caddr);
      \draw (cust) -- (cdemo);
      \draw (hdemo) -- (iband);
    \end{tikzpicture}
  }
  \caption{Join Trees used in experiments for the Retailer, Favorita, Yelp and
    TPC-DS datasets.}\label{fig:join-trees}
\end{figure*}
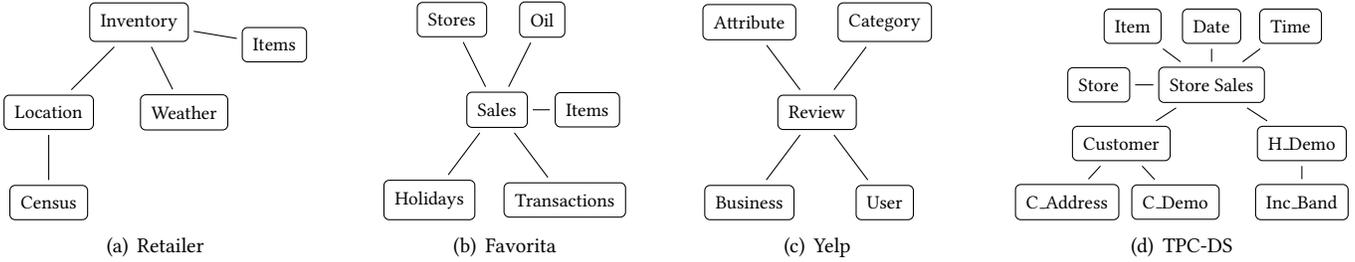

\section{Datasets}\label{appendix:datasets}

Figure~\ref{fig:join-trees} gives the join trees for the four datasets used in
the experiments in Section~\ref{sec:experiments}.

{\bf Retailer} has five relations: \textit{Inventory} stores the number of
inventory units for each date, store, and stock keeping unit (sku);
\textit{Location} keeps for each store: its zipcode, the distance to
competitors, the store type; \textit{Census} provides 14 attributes that
describe the demographics of each zipcode, including the population or median
age; \textit{Weather} stores statistics about the weather condition for each
date and store, including the temperature or if it rained; \textit{Items} keeps
the price, category, subcategory, and category cluster of each sku. 

{\bf Favorita} has six relations. Its schema is given in
Figure~\ref{fig:join-tree-favorita}. \textit{Sales} stores the number of units
sold for each store, date, and item, and whether the item was on promotion;
\textit{Items} provides information about the skus, such as the item class and
price; \textit{Stores} keeps information on stores, like the city they are
located it; \textit{Transactions} stores the number of transactions for each
date and store; \textit{Oil} provides the oil price for each date; and
\textit{Holiday} indicates whether a given date is a holiday.

{\bf Yelp} has five relations: \textit{Review} gives the rating and date for
each review by users of businesses; \textit{User} keeps information about the
users, including how many reviews they made, or when they joined;
\textit{Business} provides information about the business, e.g., their location
and average rating; \textit{Category} and \textit{Attribute} keep the
categories, e.g., Restaurant, and respectively attributes, e.g., open late, of
the businesses. A business can have many attributes and categories.

{\bf TPC-DS}~\cite{tpcds} is an excerpt of the snowflake query with the Store
Sales fact table and scale factor 10. We consider the ten relations and schema
shown in Figure~\ref{fig:tpcds:schema}. We modified the generated relations by
(1) turning strings into integer ids, (2) populating null values, and (3)
dropping attributes that are not relevant for our analytics workloads, e.g.\@
street name or categorical attributes with only one category. We provide further
details on the modifications and the scripts to do them on our website: \url{https://github.com/fdbresearch/fbench/tree/master/data/tpc-ds}.

{\bf Test Data} In order to assess the accuracy of a learned linear regression
model, we separate test data for each dataset that the model is not trained
over. The test data constitutes the sales in the last month in the dataset, for
Retailer and Favorita, and the last 15 days for TPC-DS. This simulates the
realistic usecase where the ML model predicts future sales.

\section{Experimental Evaluation}\label{appendix:experiments}

{\bf\noindent Experimental Setup} Since DBX is only available in the cloud, we
run all experiments in Section~\ref{sec:aggcomp} on a dedicated AWS d2.xlarge
instance with Ubuntu 18.04 and four vCPUs.

The experiments in Section~\ref{sec:exp3} are performed on an Intel(R) Core(TM)
i7-4770 3.40GHz/64bit/32GB with Linux 3.13.0/g++6.4.0 and eight cores.

We used the O3 compiler optimization flag and report wall-clock times by running
each system once and then reporting the average of four subsequent runs with
warm cache. We do not report the times to load the database into memory.  All
relations are given sorted by their join attributes.

{\bf\noindent Setup for Aggregate Computation}
The covar matrix and regression tree node are computed over all attributes
in Yelp, all but the join keys in Retailer and TPC-DS, and all but date and item
in Favorita. We compute all pairwise mutual information aggregates over nine
attributes for Retailer, 15 for Favorita, 11 for Yelp, and 19 for
TPC-DS. These attributes include all categorical and some discrete continuous
attributes in each dataset.  For data cubes, we used three dimensions and five
measures for all experiments. We provide DBX and MonetDB with the same list of
queries as LMFAO, which may have multiple aggregates per query.

In Figure~\ref{fig:lmfaobreakdown}, the baseline is computed with
AC/DC~\cite{ANNOS:DEEM:18}, which is a (imperfect) proxy for computing the
covar matrix in an interpreted version of LMFAO without optimizations.

\nop{
}


{\bf\noindent Setup for Model Training} We learn linear regression and regression tree
models over all attributes but join keys for Retailer, and all but date and item
for Favorita. For TPC-DS, we learn classification trees over all attributes but
join keys.

LMFAO and AC/DC first compute the covar matrix and then optimize the parameters
over it using gradient descent with Armijo backtracking line search and
Barzilai-Borwein step size~\cite{ANNOS:DEEM:18}. MADlib computes the closed form
solution of the model with ordinary least squares over the non-materiali\-zed view
of the training dataset. (OLS is the fastest approach supported by MADlib for
this problem.) We evaluate the accuracy of the model by computing the
root-mean-square-error over the test dataset and by ensuring that it is the same for
both LMFAO's model and MADlib's closed form solution.

TensorFlow requires as input the materialized training dataset shuffled in
random order. TensorFlow fails to shuffle the entire dataset and runs
out-of-memory during learning when the entire dataset is represented and
shuffled in-memory with Python Pandas. We therefore materialize and shuffle the
datasets in PSQL, and use TensforFlow's iterator interface to load the
dataset. The model is then learned with the default settings of the built-in
LinearRegressor Estimator, which optimizes the parameters with a variant of
stochastic gradient descent called FTRL~\cite{ftrl}. We used a batch size of
500K for learning, because this gave us the best performance/accuracy tradeoff
amongst all batch sizes we considered. We could not set TensorFlow to run until
convergence easily, so we computed the time it takes for one epoch (one pass
over the training data) and compared the accuracy of the obtained model with the
closed form solution.

All systems learn the decision trees with the CART algorithm~\cite{cart84}. As
cost function, we use the variance for regression trees and the Gini index for
classification trees. The maximum depth of the tree is 4 (i.e.\@ at most 31
nodes), and minimum number of instances to split a node is 1000. Continuous
attributes are bucketized into 20 buckets. We verify that LMFAO learns decision
trees that have the same accuracy as the decision trees learned in MADlib.

We used TensorFlow's built-in BoostedTrees Estimator with a batch size of
1M to learn decision trees. Larger batch sizes cause either out-of-memory errors
or a lot of memory swaps, which significantly degrade the performance. For
continuous attributes, TensorFlow requires the buckets as input, and we provide
it with the same buckets as LMFAO.

We used PSQL to compute, shuffle, and export the join results.  We tuned
PSQL for in-memory processing by setting its working memory to 28GB, shared
buffers to 128MB, and turning off the parameters fsync, synchronous commit, full
page writes, and bgwriter LRU maxpages.


{\noindent\bf Limitations of Competitors} We detail here further
limitations of the systems we encountered while preparing the experiments.  (1)
The iterator interface of TensorFlow is both a blessing, because it allows
TensorFlow to compute models over large datasets, but also a curse, because of
its overhead and poor performance for learning models over large datasets.  In
particular, it needs to repeatedly load, parse and cast the batches of tuples.
(2) R can load at most 2.5 billion values, which is less than the training
datasets require.  (3) Scikit-learn and StatsModels succeed in loading the
training dataset, but run out of memory during the one-hot encoding.  (4)
Scikit-learn and StatsModels require that all values have the same type, so they
go for the most general type: Floats. This can add significant overhead and is
one of the reasons why the Python variants run out of memory.
(5)
We attempted to benchmark
against EmptyHeaded~\cite{emptyheaded}, which computes single aggregates over
join trees. It, however, requires an extensive preprocessing of
the dataset to turn the relations into a specific input format. 
This preprocessing step introduces significant overhead, which, when
applied to our datasets, blows up the size of the data to the extent that it no
longer fits into memory. For instance, the Inventory relation in the Retailer
dataset (2GB) is blown up to more than 300GB during preprocessing. Our
observation is that EmptyHeaded has difficulty preprocessing relations whose
arity is beyond 2.  We were therefore unable to compare against EmptyHeaded.


\begin{figure}
  \subfigure[\label{fig:aggarray}Aggregates are stored and accessed
  consecutively in fixed size array.]{
    \begin{minipage}{8.5cm}\small
      \texttt{aggregate[435] += aggregatesV3[0];}\\[-0.3em]
      \hspace*{4em}\texttt{...}\\
      \texttt{aggregate[444] += aggregatesV3[9];}\\
      \texttt{aggregate[445] += aggregate[0]*aggregatesV3[0];}\\[-0.3em]
      \hspace*{4em}\texttt{...}\\
      \texttt{aggregate[473] += aggregate[28]*aggregatesV3[0];}\\
      \texttt{aggregate[474] += aggregate[0]*aggregatesV3[1];}\\[-0.3em]
      \hspace*{4em}\texttt{...}\\
      \texttt{aggregate[476] += aggregate[0]*aggregatesV3[3];}\\
    \end{minipage}
  }
  
  \subfigure[\label{fig:loop-synthesis}Updates to consecutive aggregates are
  fused into tight loops.]{
    \begin{minipage}{8cm}\small
      \texttt{for (size\_t i = 0; i < 10;++i)}\\
      \hspace*{1em}\texttt{aggregate[435+i] += aggregatesV3[i];}\\
      \texttt{for (size\_t i = 0; i < 29;++i)}\\
      \hspace*{1em}\texttt{aggregate[445+i] += aggregate[i]*aggregatesV3[0];}\\
      \texttt{for (size\_t i = 0; i < 3;++i)}\\
      \hspace*{1em}\texttt{aggregate[474+i] += aggregate[0]*aggregatesV3[i];}\\
    \end{minipage}
  }
  \caption{Snippet of code generated by LMFAO that shows how aggregates are
    stored and computed.}
  \label{fig:aggarray-with-loop-synthesis}
\end{figure}

\section{LMFAO Compilation}\label{appendix:compilation}

Recent work uses code compilation to reduce the interpretation overhead of query
evaluation~\cite{NeumannCompilation, KerstenVLDB18,
  ShaikhhaSigmod16,Shaikhha:2018, Voodoo, TahboubSigmod18}. The compilation
approach in LMFAO is closest in spirit to DBLAB \cite{ShaikhhaSigmod16}, which
advocates for the use of intermediate representations (IR) to enable code
optimizations that cannot be achieved by conventional query optimizers or query
compilation techniques without IRs.

The various optimization layers of LMFAO can be viewed as optimizations over the
following increasingly more granular IRs: (1) the join tree; (2) orders of join
attributes; and (3) the multi-output optimization that registers the computation of
aggregates at specific attributes in the attribute order.

LMFAO relies on these IRs to identify optimizations that are not available in
conventional query processing. For instance, the join tree is used to identify
views that can be grouped and evaluated together as one main computational unit in LMFAO (c.f. Section~\ref{sec:merging-views}).
A view group works on a large amount of data at once.
This departs from standard query processing that pipelines 
tuples between relational operators in the execution plan for one query. 

The three IRs exploit information about the workload at compile time to specialize the generated code. We next explain some of these optimizations; further code optimizations have been
already mentioned at the end of Section~\ref{sec:multi-output}.

\smallskip

{\em Code Specialization} The database catalog and join tree provide a lot of
statistics that LMFAO exploits to generate specific data structures to represent
the relations. For instance, given its size and schema, a relation is represented
as a fixed size array of tuples that are represented using specialized C++
structs with the exact type for each attribute. For each join, the attribute order
gives the join attribute and the views that are joined
over. LMFAO uses this information to generate specialized code that computes
these joins without dynamic casting and iterator function calls.

\smallskip

{\em Fixed size arrays} The registration of aggregates to the attribute order
allows us to derive at compile time how many aggregates are computed
in a group and the order in which they are accessed. LMFAO uses this information
during the code generation to generate fixed size arrays that store all
aggregates consecutively, in an order that allows for sequential reads and
writes. Figure~\ref{fig:aggarray} presents a snippet of generated code that
computes partial aggregates for the covar matrix. Each aggregate array is
accessed sequentially, which improves the cache locality of the
generated code.

\smallskip

{\em Loop Synthesis} The sequential access to the array of aggregates further
allows us to compress long sequences of arithmetic operations over aggregates
addressed in lockstep into tight loops, as shown in
Figure~\ref{fig:loop-synthesis}. This optimization allows the compiler to
vectorize the computation of the loop and reduces the amount of code to compile.

\smallskip

{\em Avoid repeated computation of UDAFs} As presented in
Section~\ref{sec:multi-output}, the MOO decomposes the computation of aggregates
over the attribute order. This allows LMFAO to register functions to the lowest
possible node in the attribute order. The effect of this is that we evaluate
each function only when necessary, and we minimize the number of updates to each
aggregate. For instance, in Figure~\ref{fig:mooexample} the function $g(item)$
is evaluated only once per item value, and not repeatedly for different dates
and stores that are joined with this item value.

\smallskip

{\em Inlining Function Calls} LMFAO knows which UDAFs are computed at compile
time, and can thus inline them during code generation. For instance, LMFAO
generates the following code snippet for a product of three functions that
constitutes one decision tree node aggregate over Retailer:
{\small
\begin{align*}
  \mathtt{aggregate[96] = }
  &\mathtt{(t.avghhi <= 52775\; ?\; 1.0 : 0.0)}\\
  &\mathtt{{}*(t.area\_sq\_ft > 93580\; ?\; 1.0 : 0.0)}\\
  &\mathtt{{}*(t.distance\_comp <= 5.36\; ?\; 1.0 : 0.0);}
\end{align*}
}
An interpreted version of LMFAO would make one function call for each term in
the product to compute this aggregate.

\smallskip

{\em Dynamic Functions} Some workloads require repeated computation of slightly
different aggregates. To learn decision trees, for instance, we repeatedly
compute the same set of aggregates, where the only difference is one additional
threshold function per node. To avoid recompiling the entire generated code for
each decision tree node, we generate dynamic functions in a separate C++ file
with a few lines of code. This file can be recompiled efficiently and
dynamically loaded into a running instance of LMFAO to recompute the aggregates.

\nop{

}

\end{document}